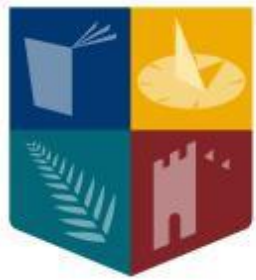

# Computational Thinking in Education: Where does it fit?
A systematic literary review

*James Lockwood, Aidan Mooney[1]*
james.Lockwood@nuim.ie, aidan.mooney@nuim.ie
Department of Computer Science,
Maynooth University,
Maynooth, Co. Kildare,
Ireland

Date: March 2017

# Abstract
Computational Thinking (CT) has been described as an essential skill which everyone should learn and can therefore include in their skill set. Seymour Papert [1] is credited as concretising Computational Thinking in 1980 but since Wing [2] popularised the term in 2006 and brought it to the international community's attention, more and more research has been conducted on CT in education. The aim of this systematic literary review is to give educators and education researchers an overview of what work has been carried out in the domain, as well as potential gaps and opportunities that still exist.

Overall it was found in this review that, although there is a lot of work currently being done around the world in many different educational contexts, the work relating to CT is still in its infancy. Along with the need to create an agreed-upon definition of CT lots of countries are still in the process of, or have not yet started, introducing CT into curriculums in all levels of education. It was also found that Computer Science/Computing, which could be the most obvious place to teach CT, has yet to become a mainstream subject in some countries, although this is improving. Of encouragement to educators is the wealth of tools and resources being developed to help teach CT as well as more and more work relating to curriculum development. For those teachers looking to incorporate CT into their schools or classes then there are bountiful options which include programming, hands-on exercises and more. The need for more detailed lesson plans and curriculum structure however, is something that could be of benefit to teachers.

---

[1] Corresponding author

# Table of Contents

# 1 Introduction

In the Republic of Ireland, as with lots of other countries, Computer Science (CS) or Computing is not yet a subject that students can sit a state exam in. Although steps to include it have been taken, so far, all that is available to students in the curriculum is a coding short course [3]. Although programming is a very useful skill and one that can be beneficial to students in a wide variety of careers and paths in life, it is not the only part of CS. Lu et al. [4] compare programming in CS to a literary analysis in English or proof construction in mathematics; it is a more advanced tool than say just being able to read and write. Jeannette Wing (2003) in her seminal paper [2] outlined how she believed Computational Thinking (CT) represents *"a universally applicable attitude and skill set everyone…would be eager to learn and use"*. She believes all children should be taught CT, placing it alongside reading, writing and arithmetic in terms of its importance. Although academics since then have failed to agree on a universal definition she states that it *"involves solving problems, designing systems, and understanding human behavior, by drawing on the concepts fundamental to computer science."* She states that it is not programming and that it means "more than being able to program a computer. It requires thinking at multiple levels of abstraction". In 2008 Wing [5] gave a further description of CT. She discussed how CT is influencing research across disciplines and that it is a skill that is being and should be used and taught to everyone. She gives two visions which are as follows:

- CT will be instrumental to new discovery and innovation in all fields of endeavour
- CT will be an integral part of childhood education

She also poses questions to CS, learning sciences and education communities with one question being: What are effective ways of learning (teaching) CT by (to) children? This in turn raises further questions about what concepts to teach, the order and using tools to teach it. She also discusses technology, societal and science drivers towards CT.

Since Wing's first publication a lot of work has been done across the world and in all levels of education to try and include CT into schools, colleges, afterschool clubs, etc. This has mainly been done through Computer Science or Computing classes/courses. As CT is vitally important to a Computer Scientist this makes sense, however, it should be noted that from the outset it is generally agreed that being able to think computationally, which includes skills such as decomposition, abstraction, algorithmic thinking and pattern matching, is of huge benefit to all disciplines.

As part of this movement researchers at the Department of Computer Science, Maynooth University, Ireland, designed the PACT program. PACT is an acronym for Programming ^ Algorithms = Computational Thinking. The hope was to introduce Irish secondary school students, and teachers, to Computer Science through both Programming and Algorithms, with the hopes of improving the vital skill of CT in participating students. Now in its 4$^{th}$ year, the PACT program has been delivered in over 60 schools and to over 1000 students.

With the introduction of programming into the curriculum and the call from administrators and governments to include more CS content in schools, the hope of the group is to develop a course which will teach CS topics whilst focussing on teaching CT. Recently, the Irish government have fast-tracked plans to introduce a Computer Science curriculum in the last two years of Irish secondary school education (http://www.education.ie/en/Publications/Corporate-Reports/Strategy-Statement/Action-Plan-for-Education-2017.pdf). It is hoped that CT will be incorporated in to this new curriculum. To this end a systematic literary review has been carried out to see how other countries and institutions have included CT into their contexts.

The hope is that this review will be of benefit to other educators, researchers, teachers and industry members who hope to introduce this vital 21$^{st}$ century skill to the next generation.

## 1.1 Note on education terms:

Primary/elementary school usually refers to the first stage of compulsory education. This usually begins at 4-5 years of age. It continues then for 5-8 years and usually finishes when students are 11-12 years old.

K5 is usually used to refer to the whole of the primary school system; this can differ from country to country but is usually from 4/5/6 years old to 11/12/13 years old.

K9 is used to refer to a mind-way point in Secondary school or the end of Middle school in countries where this applies. It usually ends around age 14/15/16. K9 sometimes concludes with a state-wide exam such as Junior Certificate in Rep. of Ireland and GCSE's in the UK.

K12 refers to the whole primary and secondary education and usually ends at age 16/17/18, whenever final state exams are given or compulsory education ends.

High school/secondary school refers to post-primary education. The age range differs between countries and sometimes even between states/counties but in general it can be seen as being from 11/12 - 17/18 years old. It usually culminates with state/country-wide exams. Secondary school is compulsory in almost all countries and although the upper age limit differs from country to country it is usually between 16-18 years old.

Some countries, such as the USA, have middle school. This is between primary school and high school and is usually equivalent to the first years of secondary school in countries where it's not found. The age range is usually 11/12 - 14/15.

College/University/Higher Education etc. all refer to tertiary education that is usually not undertaken until a student has completed primary and secondary school and is usually begun as an adult. In different countries college and similar terms may refer to different institutions with different entry criteria, awarding standards etc.

## 2 Research questions

The first stage of this review was to develop research questions. To this end, several research questions were defined to cover the potential studies involved in looking at Computational Thinking (CT) in schools. The defined questions are as follows:

- What methods/tests/tools exist to test students computational thinking ability/improvement?
- Why is Computational Thinking important for educational institutions to incorporate into their curriculums? i.e. What benefits does CT have?
- How has Computational Thinking been incorporated into already existing subjects/courses?
- What tools have been developed/used to teach CT?
- How has CT been taught in educational institutions?

## 3 Method

### 3.1 Introduction

This systematic literature review is based on Kitchenham's method [6] as applied to software engineering. This method of performing a review was chosen as the process is well documented and is derived from review processes that were previously well established in the medical community.

The paper outlines how to identify the need for the review, how to develop a strict protocol to follow for the review and how to report the findings from the review.

### 3.2 Search terms

In this study one primary search term was used, this was "Computational Thinking".

Secondary terms were used along with this and were as follows:

school, secondary school, high school, post-primary school, middle school, second level, primary school, elementary school, junior school, university, college, Ireland, third level, higher education, benefits, assessment, test.

## 3.3 Resources searched

An extensive search of the following databases was carried out between October and November 2016 using the search terms introduced above:

- ACM Digital Library
- IEEE Xplore
- ERIC
- Google Scholar

The ACM Digital Library (ACM DL) contains over 430,000 full text papers. Using the primary search term in conjunction with all the secondary resulted in 363 unique papers being returned. After eliminating by title, removing those that were clearly not related, 160 papers were kept.

The IEEE Xplore database contains three million citations. Using the primary search term in conjunction with all the secondary resulted in 107 unique papers being returned (not including one's that were found in ACM). After eliminating by title, removing those that were clearly not related, 63 papers were kept.

The ERIC database was then searched as the database is specifically for papers relating to education. We used the same search criteria as the ACM DL and IEEE Xplore and this search returned 9 papers not found in either ACM or IEEE. After eliminating by title, removing those that were clearly not related, 7 papers were kept.

Using the software tool Publish or Perish the first 1000 entries in a Google Scholar search (which combined all secondary search terms with Computational Thinking) resulted in 882 papers when compared to the search results for previous searches. After eliminating by title and removing those that were clearly not related, 136 papers were kept.

## 3.4 Document selection

Inclusion and exclusion criteria were developed as recommended in the systematic review guidelines [6]. Texts were included that:

- Directly answered one or more research question
- Were related to the teaching of Computational Thinking in educational institutions
- Were about computing or Computer Science courses/modules etc. that had a focus on problem solving/analytical skills (i.e. didn't implicitly teach CT but included relevant steps)

Studies were excluded that:

- Were in the form of a book or grey literature (opinion pieces, technical reports, blogs, presentations etc.)
- Were not published in reputable (i.e. peer-reviewed) sources
- Did not answer any research questions
- Were not published in English

After these steps, which involved reading the abstracts and eliminating papers if they did not satisfy the above criteria, 210 papers were left.

## 3.5 Quality Assessment

The following questions were taken from Kitchemann's framework [6] and each paper was analysed based off them. It was found that 35 papers were rejected based on them and the rest were all deemed eligible.

- How well does the evaluation address its original aims and purpose?
- How well was data collection carried out?
- How clear and coherent is the reporting?

- How has knowledge or understanding been extended by the research?

# 4 Related Work

There have been several literary style reviews and overview's written on the current state of CT in schools in specific countries as well as suggestions for frameworks for it. The following section presents several of these which were found during the search process.

Qualls & Sherrell [7] presents a brief overview of some examples of how CT has been used and gives evidence as to why CT should be integrated into the curriculum. In the study, some examples are given of CS courses that are adapted to introduce CT, interdisciplinary attempts to integrate CT as well as standalone CT college courses are presented. The authors discuss the Alice programming language and how it is a popular choice in college courses as well as describing the Computer Science for High Schools program developed by Carnegie Mellon University, one of the first major attempts to spread CT.

Schulte et al. [8] present an overview of a working-group report on CS at school conducted in 2012. This was an international effort which included a workshop at the Koli Calling conference in 2011 and an online survey which was analysed. They emailed the survey to known experts as well as a variety of CS education mailing lists. They received 84 respondents from 22 countries and included a variety of institutional backgrounds including school, industry and university. Based on an initial survey of the results they used a SWOT (Strengths, Weaknesses, Opportunities and Threats) inspired analysis and some of their findings were:

- CS is most often available in upper secondary school
- In upper secondary school the most relevant topics were rated as:
    - Introductory programming (rated much higher than all other topics)
    - Algorithms
    - Advanced programming
    - Programming project
    - HTML
- For upper secondary school the top goals of CS were:
    - Developing thinking skills
    - Developing problems solving skills
    - Learning programming
    - Improving algorithmic thinking
    - Databases: design and queries
- For primary school, most topics weren't seen as important with applications being the only moderately important one followed by ethics & privacy.
- For all levels of teaching, the following teaching methods are listed in order of importance:
    - Classroom based
    - Using standard applications like Microsoft Word
    - Email
    - Individual and small group work
    - Programming, projects
- Suggested problems related to CS were:
    - A lack of trained teachers
    - The perception that CS is the same as ICT
- Trends with CS:
    - Many noted that new CS curricula were coming in in the next 5-10 years
    - There is more CS in high school and an increasing demand for it

- Teaching education goals:
  - Basic concepts of CS
  - CS education goals, Education/Pedagogy, CS specific teaching approaches, the nature of CS
- Problems with CS teacher education:
  - Lack of existing teachers in the subject
  - Very little teacher education to teach CS in schools
  - No CS in curriculum
- Trends with CS teacher education:
  - Technology will evolve - This includes an increase in the variety of ways computers are used at schools, and the new ways of development of learning materials
  - CS becoming recognised
  - Teachers being trained in CS

Lye & Koh [9] attempted to answer the following questions through a review of literature:

- How has programming been incorporated into K-12 curricula?
- What are the reported outcomes of student performance in CT dimensions?
- What intervention approaches are being used to foster CT?

In the paper, they explain their view of what CT is and their search strategy and present the following findings:

- Most programming languages used were low-floor (easy to pick up) and high-ceiling (allow sophistication)
- Students used programming to learn languages and maths
- Programming taught concepts such as variables and loops
- In higher education students who used things such as pair programming and mind-mapping performed well
- Visualisation output of programming helps students learn CT
- The following are suggested interventions to teach CT:
  - Reflection
  - Constructing programs with scaffolding
  - Reinforcement of computational concepts
  - Information processing

Lye & Koh suggest the following research gaps:

- Explore more class-based interventions
- Explore more studies in computational practices and computation perspectives
- Examine the programming process

They also give some feedback on some instructional implications of the literature, such as, having a scaffolding process and giving students an authentic problem.

Mannila et al. [10] discuss the state of CS in education, in 2014, at the lowest level of education in multiple countries. They discuss both formal education and out of school CS teaching and most data was collected from a survey distributed to K-9 teachers. Some conclusions they make include:

- The inclusion of CT aspects in the curriculum is relevant in all countries
- CT concepts can be taught through various subjects
- Proper teaching education seems to be one of the most crucial factors in CS
- Programming in one form or another seems to be key to the future of CS education
- There are many positives to informal activities, however, they are not usually integrated into curriculum so students may not see the connections of CT concepts, there's also no consensus on what should be taught

Nine hundred and sixty-one people filled out their survey on teacher's experience of CT and how they felt they were teaching different topics such as algorithms, abstraction etc. They also give some examples of activities and lesson plans which could help to integrate CT into education.

Duncan & Bell [11] discuss courses developed in England, Australia and the CSTA (Computer Science Teachers Association) curricula for teaching CS to primary school before discussing their own study. They note that CS curriculum need not be developed from scratch and that others do exist. They then summarise topics found in the curricula which are:

- Algorithms
- Programming
- Data representation
- Digital devices and infrastructure
- Digital applications
- Human & computers

They also provide details of when and how the topics are introduced in the respective curricula. Subsequently they give an overview of their planned course and how they plan to assess it as well as some initial feedback from a pilot study.

In regards to the CSTA curriculum mentioned, an overview of the curriculum was published by Tucker et al. [12] following an ACM working group. In it they give an overview of the learning goals of the curriculum as well as giving sample activities that implement these. This however is not the latest version of the curriculum, the latest being detailed at https://www.csteachers.org/resource/resmgr/Docs/Standards/2016StandardsRevision/INTERIM_StandardsFINAL_07222.pdf.

Garneli et al. [13] present a systematic literature review of computing education in K-12 schools. They were specifically looking at questions relating to the benefits & challenges of using programming tools in K-12, the contexts for improving students learning in K-12 computing education and what the most common instructional practices are. Subsequently they then give a summary of what they found. Some interesting findings were:

- Thirty-three different computing tools were found and that making the decision as to which they will use is an important but not simple decision
- Game design is popular when introducing students to computing (in line with the findings of this review)
- Tangible construction kits and robotics are also popular
- The educational context/instructional method should be carefully chosen but is very hard to find classes where features are similar so that conclusions from the literature are hard to come by

Barcelos & Silveria [14] discuss how CT and Maths are related and how three different skill areas can be developed by both. They extract this information based on Brazilian and Chilean curriculum guidelines. The areas are:

- Mathematical representations and algorithms
- Establishing relationships and identifying pattern regularities
- Descriptive and representative models

They conclude that this is just one example of how CT can be introduced into curriculums without a specific CS course.

Kalelioglu et al. [15] present a literature review through which they hoped to examine the definition, scope and theoretical basis of CT. They found that a large percentage of the papers (43/125) reviewed were to do with the integration of CT into education and discussions of courses and activities based on CT. They found that there is no accepted or well-known definition of CT and that the research is in its early stages. They found that the top 5 skills related to CT are:

- Abstraction
- Algorithmic thinking

- Problem solving
- Pattern recognition
- Design-based thinking

They then present a framework for CT as a problem-solving process.

Liu et al. [16] present an overview of a series of K-12 teacher training programs that target improving CS education. Programs discussed are as follows:

- Georgia Computes at Georgia Institute of Technology
- Disciplinary Commons Program at Georgia Institute of Technology
- Summer Institute for Advanced Placement CS teachers at University of California, Los Angeles
- Linking Mathematics and CS at Purdue University
- A bioinformatics workshop at Winona State University
- Teacher Enrichment in CS course at Saint Joseph's University
- CS4HS at Carnegie Mellon University
- CSBots at Carnegie Mellon University
- An Alice training course at Duke University
- Computational Thinking for the Sciences at Marquette University
- Computing Institute for K-12 Teachers program at Lamar University

The authors then compare the above courses with each other using categories such as length, subject, grade and organiser as well as training tools used. The paper concludes that the training programs have been shown to be an effective means of making CS accessible to teachers. The authors also note that although there is currently not enough evidence to show that the same impact was made on students, assessment results of those training programs showed that most teachers could incorporate knowledge and materials gained from the training into their classes.

Voogt et al. [17] discuss CT, how it can be addressed in both formal and informal education as well as giving research and practice agendas. After giving an overview of what CT is based on Wing and Papert the authors give a more in-depth look into research history in relation to CT and its implications. They then discuss challenges in defining CT and give an overview of different definitions given in research papers. Following this they then move on to discuss CT in education including the rationales for including it and what should be taught. They then give examples of where CT curricula have been implemented, including CT as a separate subject and integrated with other subjects. They conclude that although work has been done, more is needed in relation to integrating CT in education and integrating a CT curriculum.

Gross et al. [18] reported on several lab and class attempts to introduce CT concepts to engineering students. One such course was in RWTH Aachen University in which they taught students using MATLAB and Lego Mindstorms NXT. It was designed to help teach the mathematical basis of problems, overall students reported that they learned a lot during the course and that they could apply what they learned. Virginia Tech designed a course on Computational Methods which contained modules on programming, numerical method algorithms and CT which itself included iterations, ordering and problem-solving. They also used MATLAB to teach these concepts, especially the numerical method algorithms. University College London (UCL) use a scenario-based learning technique where electrical engineering students are given week-long, realistic engineering projects to complete in teams. This teaches them both the practical skills of whatever their project is but also the ability to develop algorithms, validate projects and then present it to their classmates and reviewers. The paper also reports on some best practices found from the above three curricula which were:

- Introduce tools and processes that first year students can leverage throughout their studies
- Show student's practical implications early on
- Teachers should reuse the same technology in multiple courses
- Faculty should coordinate software use throughout an entire departments curriculum

- Integrate tools thoroughly throughout curricula for all applicable departments
- Ask students to represent their work

# 5 Research Question 1: What methods/tests/tools exist to test students computational thinking ability/improvement?

## 5.1 Introduction

As with any fundamental skill we teach children (reading, writing, arithmetic etc.) it is important to have ways of assessing and testing whether they remember and understand the skills and techniques they have learnt. It is also important that educators can assess their ability to adapt and use those skills for themselves, that they understand the concepts and it isn't just a memory exercise. The same is true for CT and without methods of assessing students understanding and ability at various CT concepts then it is hard for educators, researchers, students and administration to say whether a particularly teaching method, program, curriculum etc. is working effectively. In this section, we will look at a variety of tools and ways in which people have assessed CT concepts, assessed student's views of CT/CS as well as how different tools can be used to "grade" CT exercises.

Scratch (https://scratch.mit.edu/) has been the focus of several models/tools to assess how much CT is employed through developing Scratch projects. Seiter & Foreman [19] present a model called PECT (Progression of Early CT) and they break PECT into three fundamental components:

1) CT concepts
2) Design Patter Variables (DPV's)
3) Evidence Variables (EV's)

In Scratch the EV's are the code blocks themselves and they are given values depending on the complexity of what they do. DPV's refer to the ability to recognise the need for and use a specific design pattern for a specific problem and to use commands, syntax, variables etc. within the chosen design pattern. These DPV's are based around common coding patterns in Scratch such as "Conversate" (getting sprites to talk) and is then ranked on how advanced the usage is, i.e. one sprite speaking or having two respond to each other. Preliminary results showed that CT skills increased in projects as the creator got older, which would be expected.

Dr Scratch (http://www.drscratch.org/) is an online tool which can be used to assess CT skills through several metrics and Moreno-leon et. al. [20] compared its assessment with McCabe's Cyclomatic Complexity [21] and Halstead's metrics [22]. These are classic software engineering metrics that are globally recognised as a valid measurement for the complexity of a software system. Moreno-leon et al. found positive and significant, moderate to strong correlations and this could be considered a validation of the complexity assessment process of Dr. Scratch. However, the authors note this is a first step. One flaw of Dr. Scratch seems to be that the score of the CT dimensions measured by Dr. Scratch among several types of Scratch problems differ greatly [23]. For example, storytelling projects tended to score low in terms of logical thinking. Moreno-leon et al. found this as well as Hoover et. al. [200] who found that the quantitative (Dr. Scratch) and qualitative results differed greatly. Students designed games based around climate change and although Dr. Scratch gave similar scores based on its metrics, more complex game design and more realistic representations of climate change were found in each game (these were taken as indicators of higher levels of CT based on [20]).

Sherman & Martin [24] introduce a rubric for analysing "mobile computational thinking" (MCT) through App Inventor projects. They define MCT as a "superset of CT..., where the device changes location and context with its user". They claim that existing CT assessment tools don't cover these new ideas and therefore they tested their rubric in a mixed-major course that taught App Design using App Inventor. They measured 14 different properties. This comprised of 6 "general CT" concepts:

- Naming
- Procedural abstraction
- Variables
- Loops
- Conditionals
- Lists

in addition to 8 MCT concepts:

- Screen interface
- Events
- Component abstraction
- Data persistence
- Data sharing
- Public web services
- Accelerometer
- Orientation sensors & location awareness

These properties were rated on a "2-to 4-point scale, with increasing points representing more sophistication with the concept being measured". They give detailed examples of this scaling system in the paper. They then tested 45 apps from the 18 students. The projects were rated against the 14 elements with two of the authors rating each and any discrepancy being discussed by the authors.

The Scalable Game Design (SGD) group are based out of University of Colorado, Boulder, USA have developed several ways of assessing student's CT skills, specifically when using their AgentSheets/Cubes game design software (see Section 9.4.3). Basawapatna et al. [25] discuss how students can use AgentSheets to apply knowledge gained from programming games to creating science simulations. They wanted to see if students could recognise "Computational Thinking Patterns" (CTP), which they defined as "abstract programming patterns that enable agent interactions not only in games but also in science simulations". Some examples of CTP's are generation, absorption, diffusion and transportation which they describe in more detail in the paper. They talk about a previous method of seeing whether these CTP's have been used. In the previous method, a graph was used where the underlying code of the game is compared to canonical CTP's coded in the same programming language. This visual representation of the CTP's used in a game allows teachers to see where the student has a higher or lower proficiency and whether the students can use them. They then asked the question "To what extent are these CTPs accessible and useful to teachers teaching game design and students learning game design". They designed a quiz which they called the "Computational Thinking Pattern Quiz" (CTP Quiz) and administered it to students and teachers who took part in a game design summer institute to attempt to answer this question. The way the CTP Quiz worked was the students watched a series of videos that depict one or a combination of these CTP's and then are asked to identify which one's are used. There were 7 questions of this form with one final question in which a written paragraph described a predator/prey simulation and participants were asked to talk about all the CTPs they would use in the simulation.

Koh et al. [26] have built on the work of Basawapatna et al. [25] and developed a Cyberlearning tool to help teachers see which high-level concepts students have mastered and which they are struggling with as students code in real-time. The system is called REACT (Real Time Evaluation and Assessment of Computational Thinking) and it displays the CTP's that students are currently implementing. REACT also shows the CTP's students haven't used, as well as the correctness of previously implemented patterns. REACT offers a variety of visualisation tools and is designed to be used with SGD teams AgentSheets & AgentCubes software. Having tested the system on 134 projects with four different teachers they found the overwhelmingly positive reaction with each teacher planning to use it independently. Although they note further work and study is needed they believe this is a good first step for the system.

Basu et al. [27] developed measures to calculate CT levels in students' models and they also propose metrics to calculate correctness of computational and domain-specific constructs in these models. These models were developed in Computational Thinking in Simulation and Model-Building (CTSiM) which is a learning environment which combines CT with middle school science. They designed pre- and post-tests to measure both science content and CT skills. The CT skills were assessed by asking students to construct algorithms from scenarios using computational and scenario-specific primitives specified in the questions (e.g. Fish in an ecology based simulation). This tested student's abilities to interpret given abstractions to create useful algorithms as well as their understanding of CT constructs such as loops, conditionals and variables. The models students developed were evaluated by comparing them against the "expert model" for that activity, which gave a "correctness" value. They then had a measure for "incorrectness" which accounted for extra primitives used in the student's models. From these two values a vector-distance model accuracy matrix was developed to measure the difference between the models. It is based on the bag-of-words metric with each agent-procedure represented by the set of primitives they contain. They also developed an "effectiveness" measure which was the proportion of student's model edits which improved the accuracy of the model. They found this was a strong predictor of final model accuracies. They also implemented a measure which they called "consistency" which was the coefficient of determination from a linear regression on a student's model accuracy over time. They believe that these measures can help with online evaluation of students' models.

Gouws et al. [28] present a test for CT ability and compared students results in this test to their class grades. They aimed to develop a test that would investigate the role played by CT in introductory CS courses and to create a profile of student's abilities in CT and to see whether there is a link between the results and performance in introductory CS courses. To develop the test, they set up the following constraints:

1. The test should not rely on any existing knowledge of programming, but should test skills that are relevant to computational thinking;
2. The test should be suitable for new university students, and where possible not intimidate them or give them a negative view of their own abilities;
3. It should be possible to administer the test within a single session of reasonable length;
4. Questions should be obtained from a credible source with a genuine computer science 'flavour'.

They then sourced questions from the "Computer Olympiad 'Talent Search' papers" and for most of the questions the information was retrieved from the website. The Computer Olympiad is South Africa's implementation of the international Bebras contest. The 'Talent Search' question are designed as an aptitude test and so requires no specific knowledge of any programming language or paradigm. They then analysed the questions and classified them by 6 CT "classes" which were:

- Processes and Transformations
- Models and Abstractions
- Patterns and Algorithms
- Tools and Resources
- Inference and Logic
- Evaluations and Improvements

They selected 25 questions from the set with 6-11 being categorised in each class, some questions contained elements of multiple classes. The test was first administered to 83 new students on a CSC101 course in a pen and paper format to initially assess raw skills they possessed before any formal academic training. The second phase of testing was the to re-administer the test during the third term to the same group of students after having a semester of computer sciences studies. They found that CT results varied greatly in the initial and second test (varying from 4-88%) and that CT pass rates were significantly lower than the course pass rates (55.4% vs. 85.5%). They believe that because of this there is a significant need for their course to address and improve CT skills. They also carried out dependency tests on the class and CT tests and they found a varied result of the relationship between the two. Their results indicated that students who perform well in the assessment have a favourable pass rate for the class tests.

Werner et al. [29] describe the results of a performance assessment tool for measuring CT skills in Californian middle schools. They based their assessment on two studies. The first study was by Webb [30] during which both pre- and post-surveys were given when students were taught using AgentCubes as well as a study by Linn [31] who describes the Chain of Cognitive Accomplishments. In their course students were taught using Alice and students engaged with CT in a three-step progression called Use-Modify-Create [32] over 20 hours in a semester. At the end of the semester students were given up to 30 minutes to individually complete the "Fairy Assessment". They designed this "Fairy Assessment" as an Alice program which they hoped would analyse thinking algorithmically and making effective use of abstraction and modelling. The assessment involved solving three tasks which occurred during the playback of a narrative scenario in Alice. Failure in any of the three tasks did not result in the inability to complete the other two and they believed that for students to perform well they would "have to understand the narrative framework of the story underlying the program and to understand existing program instructions which create the framework". The test is described in detail in their paper. They found a large variety of results but in general feel their findings suggest that the Fairy assessment is a promising strategy for assessing CT because it is "motivating"; only 30 of the 311 didn't attempt to modify the program. They also suggest the test was successful in picking up a range of CT across students and a variety of types of CT across the three tasks. Task 2 was the most complex and this was reflected in the lowest mean scores.

## 5.2 Conclusion

Overall work in testing for CT is in it's infancy. Most of the examples presented in this section are in the early stages of development. Tools do exist such as Dr. Scratch and the tools developed by the Scalable Design Group but there is a need for more research into this area. Other forms of test are based on problem-solving and analytical thinking tests. Whilst these are potentially beneficial, if CT is to become a common skill taught in schools and universities then built-for-purpose tools and assessments might be required.

# 6 Research Question 2: Why is Computational Thinking important for educational institutions to incorporate into their curriculums? i.e. What benefits does CT have?

## 6.1 Introduction

For teachers, educational institutions and administrators to consider incorporating and introducing Computer Science and more specifically Computational Thinking (CT) into schools then it is important that we show what benefits students can gain from studying CT. As discussed in the introduction, CT is in part a problem solving and thinking process. Adding to student's abilities in this area could be of benefit to them in many areas of life and study. In the following section studies, which have found specific benefits or potential benefits of CT integration and teaching are presented.

Lishinski et al. [33] examined the relationship between CS1 student's (first year Computer Science students at third level) problem solving skills and their programming outcomes as measured by programming projects and multiple choice exams. To do this they used items from the problem-solving section of PISA (Programme for International Student Assessment). They gave both PISA tests and analysed the projects of 41 students and ended up with 28 datasets of students who completed all the programming projects and exams. They found that student's problem solving scores measured at the start of term are predictive of student outcomes, but only on the constructed-response programming assignments, not on the Multiple Choice (MC) exams. They also found that there is an association between MC grades and code-writing grades which supports other studies [34]. They also found that programming ability is hierarchical and goes beyond what can be assessed in a MC test. They also found that problem solving skills are uniquely predictive of higher level skills above and beyond the effect of lower level skills.

Haddad & Kalaani [35] showed that CT can be a good predictor of a student's academic success. They studied data from 982 students over the course of two years who took a "Computing for Engineers" course which is designed to formally introduce students to CT in first year. They compared student's overall Grade Point Average (GPA) to two predictor variables: the students CT skills represented by their final grade in the course and the teaching style represented by the different instructors. Their performance was assessed using a set of quizzes, weekly lab reports, homework assignments and exams. They found that student's average GPA scores correlated with their CT performance. They also noted that the difference in performance was mainly due to how students perceived CT and their level of readiness. They claim if "the instruction of CT is tailored to fit student's learning styles then more students will perform better and grasp the concepts being taught more effectively". The distribution of students' GPA and course scores had almost the same mean which supported their hypothesis that CT is a valid tool to predict future academic success. They also suggest that their results could support the idea that CT could be used as an early intervention indicator to ultimately increase students' retention, progression and graduation rates within STEM disciplines.

Van Dyne & Braun's [36] paper is discussed further in Section 9.5 but one notable finding was that they found that their new CS0 course, which was designed to prepare students for CS studies by teaching them CT skills, improved analytical skills in majors and non-majors. They used the Whimbey Analytical Skills Inventory [202] to test student's analytical skills. Pre-tests with an average score of 60.3% were recorded with a range of 31.6%-89.5% while post-tests had an average score of 78.4% with a range of 51.4%-100%. This is significantly different and shows an increase of over 18%.

Oliveira et al. [37] discuss a study on their attempt to identify a qualitative link between student's ability to compute and their performance, this case in primary school. They give a list of abilities a student must use to compute: Abstraction, Calculation, Reading & Understanding and Design & Writing. They developed a test which consisted of seven questions, five of which evaluate the student's ability to abstract and calculate using Turing Machines as a basis. The final two questions assess the student's abilities to read, understand and to design and write. They collected data from 81 primary school students and found that there was significant correlation between the test scores and academic scores for all classes. They propose that it is the first step in seeing whether improving student's ability to compute can positively affect student's performance at school.

Davies [38] reports on a study carried out in which students in an introductory programming course were taught using a structured pseudocode (see paper for examples) which was "designed to highlight and facilitate algorithmic construction". This was done as the complexity of surface features of programming languages (syntax for example) can be distracting to students and prevent them from understanding how to solve the problem conceptually. The formal programming language (C++) was only introduced in the final three weeks of the course and it was found that their knowledge of the language was equal to those who were taught C++ for 13 weeks, as examined in a final exam. Students also completed anonymous surveys and one interesting finding from that was that the control group (those taught C++) rated the statement "For me the most difficult thing about writing a computer program is coming up with the basic algorithm to solve the problem" much lower. The authors believe this supports the hypothesis that students who are taught a programming language are "fooled" into thinking the language is the major source of complexity. They also found that women especially might benefit from this method of teaching, particularly in terms of their attitudes and confidence of programming.

## 6.2 Conclusion

The discussed papers have shown that teaching CT or integrating CT concepts could:

- Improve student's analytical skills
- Provide a better understanding that programming is about solving the problem not just the code
- Improve women's attitudes and confidence towards programming
- Be used as an early indicator and predictor of academic success and that CT scores correlate strongly with general academic success

However, CT and research into it are still in the early stages, therefore long-term effects as well as additional benefits still need to be researched. The above findings are encouraging and show that CT is a beneficial skill but more research is required before the extent of the impact of teaching CT can have on students is known.

# 7 Research Question 3: How has Computational Thinking been incorporated into already existing subjects/courses?

## 7.1 Introduction

For some schools or colleges, it might not always be possible to run a complete CS curriculum or similar in which CT can be taught separately. As such it is important that efforts can be made to ensure that CT is not only taught in regards to CS but that it can also be incorporated into many other areas of education. Even if it is possible to teach CT outright it is important that students are made aware and shown how their new skills and knowledge can be applied to many areas rather than just CS or even science. In this section, we summarise papers and studies that talk about how CT can be integrated into other subjects or show that CT can be useful in the teaching, learning and application of different subjects. The hope is that the studies outlined can help educators who lack the time or resources to teach CT as a standalone subject to incorporate this important skill into their classrooms.

## 7.2 Sciences

Ahamed et al. [39] discuss a three-day workshop in which they aim to emphasise the "deep connections between the natural sciences, mathematics and computer science". They held sessions on the following (links to relevant curriculum sections given in paper):

- Computational Thinking - an introduction where they explained Wing's article on CT and topics such as abstraction, algorithms etc.
- Simulations - introduced teachers to the concepts of simulations and their relevance, used Excel - one example is Newton's Law of Cooling
- Probability - used VPython to construct a simulation as well as discussing the relevance of probability in different areas
- VPython & Python - taught teachers how to use these tools
- Mathematics - cryptography using CS Unplugged
- Physics - Inclined plane, Newton's Second law, gravity etc.
- Biology - Mendelian genetics
- Chemistry - virtual screening in drug design
- Computational Science jobs - bring awareness to what jobs are out there
- Lesson plans - allowed teachers to develop lessons plans and present them

They found participants understanding of CT and its importance to their field improved as well as improvements in teaching CT and the use of computational tools.

Hambrusch et al. [40] describe the development of an "Introduction to Computational Thinking" course which was to be taken by science majors. The course was designed with the aim to lay the groundwork for CT and it was influenced by the desires of the Physics, Chemistry and Bioinformatics departments. They taught the students Python and through it taught Computational tools and models such as the Monte Carlo method, Ideal gas simulations (Physics), graphs and modelling proteins (Bioinformatics). It also included a section on the history of CS, limits of computing and the future of computing. The assessment was done through four assignments and four projects. These projects were (they give further details about these in the paper):

- Manipulating Digital Audio

- Computational Experiments on Percolation in Grids
- Simulating Physical Systems
- Analysing Protein-Protein interactions

They found that their course increased interest in taking another CS course and from feedback students liked the problem-driven format. From this experience, they suggest that Python is an excellent first language as it is used in many scientific disciplines. They also believe that the same course can be taught to all science subjects.

Swaid [41] discusses a project that aims to bring CT into STEM disciplines at Philander Smith College, Arkansas, USA. This was done through introductory sciences course which they call "Gate-keeping courses". These included Biology 1, Calculus 1 and Programming 1 and they identified seven CT elements to adopt and discuss how they are currently used in these Gate-keeping course. These seven elements are:

1) Abstraction
2) Data
3) Retrieving
4) Algorithms
5) Design
6) Evaluation
7) Visualisation

Subsequently they propose an Introduction to CT which emphasises CT concepts through hands-on learning experiences and is a two-part course: (i) Introduce students to CT concepts and applications and (ii) Introduce students to programming (in this case Java).

Sabitzer & Paster [42] present a review on their course called "COOL Informatics" where they hope to show that informatics is "cool" and can be fun and easy. They do this by using CS Unplugged style activities but also going further by using informatics concepts such as algorithms to support learning in a variety of subjects in the curriculum. The "Cool Informatics" approach has several overarching principles which are as follows (also included are some teaching/learning methods):

- Discovery - solution based learning, video tutorials, hand's-on etc.
- Cooperation - Team and group work, pair programming
- Individuality - Questioning, competence-based learning
- Activity - Hand's on, learning by doing

They have tested this approach on several pilot projects in primary level, secondary level and higher education and they give some case studies of some of these projects. Examples include Encryption, PowerPoint and Brain-based Programming. Evaluation results in different schools and university indicate that "Cool Informatics" is appreciated by teachers and students as well as being an effective way of teaching. Exercises for discovery and step-by-step learning assist with learning and understanding complex topics.

Isbell et al. [43] present the outcome of a working group that looked at defining what computing is. They discuss why this is relevant and important now and then define computing as "any purposeful activity that marries the representation of some dynamic domain with the representation of some dynamic machine that provides theoretical, empirical or practical understanding of that domain or that machine." They then discuss the practice of computing and how it can impact other areas, in this case looking at DNA. They then go on to discuss who and what should learn about computation stating that "every student should include an exposure to the ideas of modelling, abstraction and automation", whilst giving more learning outcomes for more specialised areas.

## 7.3 Biology/Life Sciences

Qin [44] talks about the development of a course to teach Bioinformatics to biology students through which students quantitative and critical thinking skills would be improved through teaching computing and CT. The course emphasises CT and hand-on learning and to do this they created a mapping between CT and bioinformatics, an example of this is biological databases (MySQL) – abstraction, protein structure - optimisation. They found the following: biology students are uncomfortable with one-on-one interaction with computers during labs so they decided to pair people up. They ensured individual learning by giving in-class quizzes, mixing of pairs and discussions. They also found that by emphasising concepts of CT that students recognised a common thinking pattern. Post surveys showed that students were positive about their experience and felt that they had improved both their computer knowledge and quantitative skills; some students decided to take programming and computing courses after the course.

Rubinstein & Chor [45] discuss how they aimed to teach concepts such as algorithms, abstraction and logical thinking to life science students. Their course requires a knowledge of programming and then focusses on "developing student's computational thinking skills". This consisted of four modules with each relating to a different biological domain. The domains were then mapped from Biological content to Computational topics to then a computational concept. A few examples are:

- Biological networks -> Graphs, Dijkstra's algorithm -> Greedy algorithms, abstraction, reduction
- Biological sequences -> Automata and regular expressions -> Pre-processing
- Systems simulation -> Boolean/Discrete networks -> Simulation, discrete maths
- Biological images -> Image processing: edge detection -> Modular design

Students solved "real-life" problems through coding in Python and they focused on practical use rather than syntax. They feel this approach of not teaching basic programming from scratch allows them to go beyond programming and allows the computational concepts to take centre stage. Although they did not do any in-depth evaluation they found that students could focus on these notions rather than programming and they found that students view of what CS is changed from being about the physical machine to being about problem solving and other CT concepts.

## 7.4 Physics

Dwyer et al. [46] is in CS Unplugged section.

## 7.5 Maths

Jenkins et al. [47] describe a design for the immersion of CT into mathematics classrooms of high schools in Alabama, USA. They provide an overview of the state of computing in Alabama and a description of the Alabama Math, Science and Technology Initiative (AMSTI). They conducted a workshop for maths education leaders in which the teachers were given problems to solve in which they had to used abstraction, generalisation and justification. They then use Python to create mini-programs to solve the problems. Results showed that it was an effective way to get teachers to use programming in maths classes and that teachers saw it as a new tool to teach mathematical reasoning. They were also eager to learn more about programming and how to integrate it into their classes.

Sysło & Kwiatkowska [48] discuss how CT concepts can be incorporated into traditional school mathematics and help enhance the learning experience. They take topics that are present in informatics (CS) high school textbooks in Poland but are absent from mathematics textbooks and show how they can contribute. Below are some examples of the topics and how they say they are linked to maths:

- Representation of numbers -> polynomials
- Reduction and composition -> Given sides of a triangle, is it a valid triangle
- Approximation -> rounding errors -> quadratic equation
- Recursion -> hard for maths, easier after teaching in CS

- Heuristic thinking -> trial and error/greedy algorithms/shortest path etc.

Freudenthal et al. [49, 50, 51] present the content and evaluation of an introductory programming course called "Media Propelled Computational Thinking" (MPCT). It is designed to "foster student intuition in, appreciation of and confidence about basic pre-calculus topics for freshmen year students". The course is made up of a series of modules on calculus based topics where students program algebra and geometry concepts.

## 7.6 English
### 7.6.1 Journalism

Wolz et al. [52, 53] present an overview of a summer school and after school program in Interactive Journalism which was designed for middle school students and teachers to develop a competency and interest in CT. Students and teachers conducted research and interviews into how to develop news stories that are presented using Scratch animations, text and video. From their data, they believe that they changed student's perceptions of programming and that students had most fun during the Scratch programming sections of the summer school. They also increased student's confidence in their ability as well as showing them that the processes of good writing and good animation design (software design) are similar. They note however that it is hard (at time of publication) to see whether the Scratch projects do demonstrate CT.

In their paper Pulimood et al. [54] discuss the impact of their collaboration between CS and journalism on CT skills. The aim was to increase interest in and motivation toward computing careers as well as formalising a model for courses that collaborate across disciplinary boundaries and with a community partner. The collaboration occurred over five semesters with courses in Software Engineering and Blogging and Social media being the first delivered. Students worked on developing a web-based system for the non-profit organisation Habitat for Humanity. Students in the CS class focused on the technical research, design and implementation, while journalism students looked for data sources and wrote content for the website. They assessed students CT skills using surveys and they found that journalism students initially rated their skills as lower than the CS students. After the course students, all rated their CT skills as higher, they found statistical significance in all measures post-course.

### 7.6.2 Writing

Howell et al. [55] present on the first phase of a cross-disciplinary project between the English and Computing departments at the University of North Florida, USA, on how to introduce computing into other subjects. They built "Concept Maps" [56] of English structures for writing to help introduce CT into the writing method. This was due to that fact that teachers of English writing often find it hard to say what they mean when they talk about writing "clearly". They give an overview of their model and give some benefits which include: for the English faculty members, it clarified concept mapping and clarified that the knowledge of concepts is implicit rather than explicit. They also state that to transfer this knowledge into the classroom students would need to be taught concept maps (modelling) and then use it in the teaching of English concepts.

### 7.6.3 Literature

Nesiba et al. [57] talk about the DISSECT (DIScover SciEnce through CT) project which aims to introduce students to CS principles by integrating CT into middle/high school science and ask the question of whether CT can be infused into humanity subjects. They believe it can be and present one approach in which CT practices are combined with composition and literature through a 12th grade English Literature course. They provide an overview of DISSECT as well as giving some examples of how they achieved this integration such as:

- Song lyrics unit - students conducted lyric analysis, poetry device data analysis, song critique and website creating
- Macbeth - using a drag and drop comic creation tool called ToonDoo [201], students had to use algorithmic thinking to create a story board of the scene

- Lord of the Flies - students created a symbol development project which took eight symbols from the novel of the same name and constructed a shared Google doc and found every instance of said symbols, this provided use of abstraction
- They also provided some extra CT units such as introduction to algorithms, internet source reliability and summarisation

To assess the course, they used a variety of tools and metrics and they discuss these in further detail in the paper. They found DISSECT students performed better in CT assessments than the control classes and performed better in a writing test (EOCE). They also found the integration was successful and that the classes demonstrated use of their CT skills such as abstraction.

DISSECT is also talked about in the Section 9.2.

## 7.7 Dance

In their paper, Daily et al. [60] present an approach to teaching computational concepts and using them through programming a 3-D character in Alice that can perform dance moves. Each week 5th and 6th graders met to learn the basics of dance (body, space, time, energy), dance choreography and Alice programming. The authors had previously made a character in Alice which could be controlled using Alice's control system and made to move around the screen and move body parts. Some feedback they got from their initial pilot study was that the dancing motion of the 3-D character wasn't great, that it should be smoother and that parts would move unnaturally. Whilst making a program that made various dances, students used "computational concepts" such as "DoTogether", "Wait", "Loop" etc. Interviews with the participants showed they thought about how to construct the commands to achieve the desired effect, used problem solving strategies and used modularisation to make smaller pieces of dances which they then strung together. Students also used testing and debugging techniques and this often occurred in parallel with students "testing" it in the real world. Also, perspectives of computing seemed to change, students seemed to enjoy the process. The authors comment that this is a first stage and that future work is needed, but that dance and programming seem to fit well.

## 7.8 General/Multiple

### 7.8.1 Framework - general

Perković et al. [61] present a framework for implementing CT in a large range of general educational course at DePaul University, Chicago, USA. The range includes courses in codes and ciphers, animation, early Russia and urban ecology and these 19 courses were revised to include CT more explicitly. They give three specific and in-depth examples of how CT is currently in modules including Scientific Enquiry: Geographic Information Systems, Arts and Literature: Introduction to Game Design and Arts and Literature: 3-D Modelling. They give a description of how it was taught, the learning goals and how it was assessed.

### 7.8.2 Interdisciplinary

Yang et al. [62] presents a course in which they try to disseminate CT and interdisciplinary cooperation through a course on gerontechnology. Gerontechnology is technology to assist in the lives of old people, especially things such as smart home technology. They designed a course which included teaching students about the ageing process (Gerontology), assistive technology, software engineering practices and design guidelines for older adults. This was a collaboration across three departments (Design, CS and Gerontology) and they were all focussed on having CT as a cross-cutting theme. Concepts such as composition and decomposition were given a specific focus in modules on interface design and assistive technology. The course showed a significant improvement in student's self-reported competency in CT although students reported "competency to engage in CT" dropped. They state this could be due to CS students who took the course not realising that there was more to learn. It also improved student's self-reported capabilities for interdisciplinary teamwork. They also found that non-CS students who took the course performed comparably to CS students, even in CS modules.

Goldberg et al. [63] present a course in which they aim to bring computing into classes that students are already taking including art, biology and mathematics. They wanted to "include CS-based activities that allow students to reason about computational concepts" and did this by meeting with students and teachers to discuss where computational concepts might be beneficial and fit the requirements of curriculum. They then give some examples of how it was included in the respective subjects:

- Art – used Photoshop, Dreamweaver and other tools to learn about vector graphics, used Arduino and other tools to teach programming, sensors etc.
- Biology – DNA sequences -> algorithms, data analysis
- Health education – taught about queries for searching for good information, graphs and networks for a unit on STI's
- Geography – computational cartography, computer visualisation
- Social and Civil society – made pace trackers using Arduino to do a project on walking in cities, used data and correlations on data

They believe the curriculum can easily be sustained and disseminated and that all teachers involved plan to continue incorporating the computational content. Preliminary results from the teachers show that their interest in computing topics was increased and that K-12 student's engagement increased. They hope that their courses will provoke interest in CS and that students will be more likely to further study it.

Shailaja & Sridaran [64] give an overview of what, why and how to teach CT. They give a list of topics which should be covered (based on ACM/IEEE model) which includes: Algorithms, Discrete Structures, Programming Fundamentals, Graphics and visual computing and more. They then give a breakdown of CT skills they believe can be learned at different ages as well as examples of how to teach them. Below is a list of the skills by age group and one example they suggest for each skill:

- K-2 (Montessori) - integrate with already existing topics
    - Visualisation - Multimedia such as an educational CD
    - Pattern recognition - work independently on software such as paint
    - Generalisation - outputs, that a monitor/printer is like a tongue/skin
- Grade 3-5
    - Decomposition - use LOGO/Scratch to learn about angles, execution of command etc.
    - Algorithmic thinking – Writing algorithms
    - Evaluation - independent learning such as puzzles, games to pursue self-learning
- Grade 6-8
    - Abstraction - changes in technology and how they affect offices and other places
    - Critical Thinking - troubleshooting
    - Computing - use software tools (BASIC/VB suggested) to appreciate what programming is by making simple programs
- Grade 9-12
    - General CT - can teach programming languages such as Python, Java. Make games, programs for maths calculations etc.

## 7.9 Conclusion

It can be seen from the presented papers that introducing CT doesn't have to be done exclusively through new courses or even through Computer Science. CT is a skill that can be used in a possibly surprising range of disciplines and can benefit students studying in any area. The ability to break down a problem and develop a manageable solution is one that all students will find useful in both their academic and work lives. These papers show how CT can be successfully taught in varying topics and subjects which can be especially helpful to educators dealing with already crammed curriculums.

# 8 Research Question 4: What tools have been developed/used to teach CT?

## 8.1 Introduction

Many tools have been used and/or developed to teach CT in many contexts, from week-long summer camps aimed at high school age students to day-long workshops for primary school students to semester-long credit awarding university courses. From the corpus of papers in this literary review over 50 different tools, software packages and educational activities were found that had been used, developed or integrated into pedagogical approaches to CT. These varied greatly from prototype devices such as Algo.Rhythm [65] and CyberPLAYce [66] to widely available and used resources such as CS Unplugged and Scratch. In this section, we discuss the most commonly found tools in the papers and discuss briefly what they are before looking at examples of how they have been used to teach CT including their impact, popularity, concepts taught as well as recommendations from the authors.

## 8.2 Scratch

Scratch (https://scratch.mit.edu/) is a visual programming language designed by MIT Media lab which was released in 2005. Scratch helps young people learn to think creatively, reason systematically, and work collaboratively through designing their own interactive stories, games, and animations. These creations and projects can then be shared on an online community which at the time of writing has over 17 million registered users and over 20 million shared projects [67]. Scratch has been used in a few different ways to teach CT concepts.

Boechler et al. [68] developed a course aimed at teaching trainee teachers about how to incorporate CT into their classes. To do this they taught students game design using Scratch. They then tested several metrics which they identified as proxies and evidence of CT skills development. These metrics were number of scripts, number of blocks, number of variables, number of child scripts and nesting complexity (average depth). They found that the Software Recognition Test was the best predictor of three out of five of these variables (these were number of blocks, number of children scripts and average depth). Also in relation to assessing CT skills in Scratch, Dasgupta et. al. [69] found that users who use Scratch's remix tool, which allows users to take projects made by others and change them, have larger repositories of programming commands. This is even true after controlling for the number of projects and amount of code shared. They also showed that exposure to CT concepts through remixing is associated with increased likelihood of using them.

Cho et al. [70] reported on a two-day workshop (Google Computer Science for High School) delivered to teachers in which they were taught CT skills and how they can benefit everyone across disciplines. They were given brief presentations on CT concepts (abstraction, automation and analysis) as applied to topics including binary search, data representation etc. The rest (and majority) of the workshop involved hands-on activities using Scratch and App Inventor that allowed teachers to develop interactive lessons for their own fields that related to some aspect of CT. These presentations and activities were found to be helpful for incorporating CT and Scratch (and App Inventor) sessions were rated very highly. Sullivan et al. [71] used a variety of tools to teach CT concepts and computer programming. These included Scratch (which they used in the programming part), CS Unplugged and Blockly. They found that participation in the course led to an increase in student's perceived ability to program a computer and an increase in computer self-efficacy. It also led to an increased awareness of career options in CS and led to them thinking less that a CS degree is for "geeks". It didn't, however, increase their perceived likelihood of pursuing CS in further education.

DISSECT [57,58, 59] is an initiative by New Mexico State University, USA to incorporate CT into 6th grade classrooms. To do this they use a variety of different techniques and tools including algorithm writing and teaching Scratch. They used Scratch to reinforce CT skills and concepts previously taught in other modules; the Scratch module was received very positively by the students. They also used it to teach the concept of iteration by making "Web Ads" and they found that it was an enjoyable module which allowed students to demonstrate their knowledge of iteration. Using pre- and post-tests they found that this model improved students CT vocabulary and definition skills. In fact, the students' scores were comparable to those of high school students and the results suggested that 6th grade students have as much capacity to learn complex concepts such as CT as their high school counterparts. Qualitative analysis also showed that students had a more thorough understanding of the CT concepts taught and they suggest that they might be able to not only recognise CT concepts but reuse them across subjects and disciplines. It also indicated that levels of engagement and learning in the 6th grade science curriculum concepts incorporated into their modules was higher and that the modules effectively increased interest, motivation and knowledge of CS.

Grover et. al. [72,73, 74] developed a curriculum called FACT which they taught to middle school students. They used Scratch as their primary teaching tool and taught topics such as loops, variables, user input, algorithms and conditionals. They found that their course showed learning gains and fostered CT skills as measured by pre- and post-test instruments borrowed from earlier work in Israel [75] as well as quizzes and Scratch assignments. They found that their course (24 hours over roughly six weeks) didn't give statistically better results than the 60-hour Israeli course given over a year, however, there are other contributing factors such as the Israeli students being picked as those "who excelled in their age group". One question on the survey that did produce significantly different results involved filling in 10 blank Scratch blocks in a script and this involved the highest level of thinking, "Evaluating and Creating" in Bloom's taxonomy [76]. They also found that students who had low prior mathematics achievement struggled with learning CS concepts and they suggest that it points to a need to build abstraction skills that maths prepares students for.

In New Zealand, the National Certificate of Education Achievement made standards available for a CS high school curriculum in 2011 and the implementation of curricula has been heavily documented by Bell et al. [11, 77, 78]. New Zealand introduced CS as a nationally assessed topic in 2011 and it is a three-year course. They use Scratch as an initial introduction to programming and concepts before moving on to Object-Orientated programming in future years. A pilot study in 2014 was carried out in which over 500 11/12-year-old students were taught Data representation concepts using CS Unplugged and then programming using Scratch. The teacher giving this course felt that an introduction to Scratch was needed before adding in programming concepts and so introducing it earlier would increase confidence. In general students liked learning Scratch the most out of the two topics with a difference between males and females in both their enjoyment of Scratch and general enjoyment of the course. They also found that the Bebras challenge has the potential to be used as a proxy to test CT skills but more work is required.

Webb & Rosson [79] developed a curriculum based on using Scratch after initially using multiple tools (Scratch, Alice and Lego Mindstorms) as students (all girls) were confused when transferring from one platform to another. They used scaffolded examples to teach students CT concepts as well as introduce them to Scratch. They found that self-reported self-efficacy for computational problem solving, abstraction, debugging and terminology increased and the students both enjoyed the course and felt they had been successful.

Imberman et al. [80] report on a Google funded workshop for the CS4HS initiative in which they used specific tools (Scratch, Lego Robots, Raspberry Pi, CS Unplugged and App Inventor) to teach teachers how to incorporate CT and CS into their schools. Of the 24 teachers that attended, Scratch was the most well-known of the tools taught, they found that teachers were more likely to use Scratch than any other tool.

## 8.3 App Inventor

MIT App Inventor (http://appinventor.mit.edu/explore/) is a beginner's introduction to programming and app creation that transforms the complex language of text-based coding into visual, drag-and-drop building blocks. The simple graphical interface grants even an inexperienced novice the ability to create a basic, fully functional app within an hour or less of starting to use the tool.

Fronza et al. [81] describe the structure of a week-long summer school they developed called MobileDev. Through this course, they use the "curiosity of students for developing mobile apps to introduce and teach CT via programming mobile applications through exercises of increasing difficulty. They used App Inventor as the practical framework for their camp. After discussing and introducing CT with real-world and applicable examples students are given a short introduction to App Inventor. Subsequently they are given a series of simple exercises of increasing difficulty following specific CT steps given to the students such as problem decomposition, data representation and algorithm writing. Special attention is given to the design of the solution. Students are then given a day-long final project which were designed using an Agile approach. Using an assessment framework based on [205] they assessed the development of three dimensions of CT (practices, concepts, perspectives) and this was done informally through questioning. They also informally found that students enjoyed the content and the wide-variety of projects lead them to believe students (of varying backgrounds) can exercise CT skills although they say in future they will administer pre-and post-surveys.

L'Heureux et al. [82] give an overview of their course which is designed to cultivate CT in information technology education at college level. They took IT-related and business centred scenarios to come up with concepts which cultivate CT concepts whilst being relevant to real-life IT issues. These concepts included logical thinking, strategising, abstract thinking, procedural thinking, optimising and iterative refinement. They used the STAIR methodology [83] to guide students through problems. The course was then built around several modules. This included developing apps using App Inventor where they were given free rein to create an idea so long as it had a certain level of complexity. They then used Alice to teach some basic programming through animated storytelling. They were again given freedom to choose so long as it had certain elements. Other modules included developing an electronic portfolio and an investigation into desktop security. CT was measured during the course using a survey called advancing the Successful IT student through enhanced Computational Thinking. They found that every question asked was answered with strongly agree or agree 50% of the time (on average) and so they concluded that students felt they were using CT to solve problems. They found that app development in particular scored highly (almost 90%); in particular, questions which lead them to suggest that the module "pushed students to clearly define what they needed to do and understand how success of their objectives would be measured. They felt the course demonstrated that it is possible to improve students CT skills through an information technology course and that it increased student's engagement and connection with the material.

Grover & Pea [84] used App Inventor to pilot their course which they designed around "Computational Discourse". They defined this as being "where learners would be introduced to ideas of CS through building competencies in CT by knowledge building discussions in concert with engaging in computationally rich activities". It is based off a framework that emphasises the importance of social interaction in the development of individual processes. The pilot study was a day-long course with seven middle school students with a mean age of 13. The day was split into two distinct sessions. The first was introducing the students to the basics of using App Inventor. As they were investigating discourse intensive pedagogy the curriculum was focussed on discussions and questions emerging from introductory exercises borrowed from a book. In the second session students worked in pairs or alone to develop an app which they came up with themselves. They found that coding for "discourse moves" in audio and video transcripts indicated a significant role in the flow of the workshop and give examples in the paper. Preliminary results showed that CS vocabulary had increased and a growth in the use of important CT elements in the context of developing the apps was observed. They felt that App Inventor was a good tool for this kind of teaching as it led to discussions as to how to solve problems and its event-driven architecture allowed students to talk about them in a way keeping with how novice programmers approach problem solving. They also note that App Inventor compares well with Scratch and Alice but offers the benefits of having something very tangible to show and allowing students to make apps which is of interest to them and which they have a lot of interaction with.

## 8.4 Games & Game Design

Games and game design are a popular way of teaching in general [85] and have been found to be an engaging and effective way of disseminating information [86, 87, 88]. This is true of teaching CS and more specifically CT. Several papers were found that talked about game design and game playing as being useful tools for teaching CT concepts. In this section, we discuss several of them.

### 8.4.1 RaBit EscApe

Apostolellis et al. [89] designed a board game aimed at 6-10 year olds which "challenges children to orient tangible, magnetized manipulatives to complete or create paths". The game is comprised of wooden pieces which are called "bits" that are magnetised and the aim is to place these together in a predefined path and help the rabbit escape from the apes. They ran an informal study in which two groups of three 8-10 year olds played the game and from this they found that scaffolding is necessary in terms of increasing the difficulty level and that further developments such as a cheat sheet could be beneficial. The children's interactions with the game encourages them that this could be a low-tech, tangible and engaging way of teaching CT.

### 8.4.2 Board Game - Pandemic

Berland & Lee [90] present how board games can and do incorporate CT concepts and may allow people to learn whilst playing collaboratively. They collected data on three groups of first year undergraduate students playing a game called Pandemic (http://www.zmangames.com/pandemic-universe.html). They recorded the sessions and then divided the audio up into individual game turns and then again into individual statements made during these turns. These segments were then coded with respect to rules or concepts being used. The concepts are conditional logic, algorithm building, debugging, simulation and distributed computation but they state that this list is not exhaustive or mutually exclusive. They found that distributed computation (which they describe as rule based action) was the most common found during all three sessions. They also give examples of these from the recordings. They believe that players had to 1) internalise a set of rules and 2) optimise behaviour and strategies based on these rules and they believe CT concepts are in play across various strategic board games and hypothesise that these games could be an important foundation which designers could intentionally use to develop CT.

### 8.4.3 Program your robot

Kazimoglu et al. [91, 92] present a "serious game" in which they aim to teach programming and computational thinking concepts. It is an Adobe Flash game called "Program your robot" in which players must help a robot escape from a series of platforms using a "solution algorithm". Players construct this "solution algorithm" by giving various commands (split into action & programming commands) to the robot. The gameplay is based on a framework they developed to map CT concepts to the game structure. For example, problem identification and decomposition is related to Problem Solving (a CT skill) and in their game this is done through helping the robot to reach the teleporter (end point). It is similar to LightBot (https://lightbot.com/flash.html) and Robozzle (http://www.robozzle.com/) but the authors claim that those games aren't designed for learning, but fun, whereas theirs is. Although lacking empirical evidence they feel that their game encompasses the following CT skills: algorithm building, conditional logic, tracking a simulation, debugging. They ran an informal test on their game with CS students who had all studied at least one programming course. The feedback from the students was that they enjoyed the game and that this type of approach could enhance the problem-solving skills of introductory programming courses.

### 8.4.4 Agentsheets/AgentCubes

AgentSheets and AgentCubes (http://www.agentsheets.com/index.html) are tools that let people create their own agent-based games and simulations and publish them on the Web through a user-friendly drag-and-drop interface. Interactive simulations help students grasp new ideas, test theories, explore complex processes in various science fields. They state that building games teaches students Computer Science concepts, logic, and algorithmic thinking. The design of AgentSheets and AgentCubes as well as studies on them are carried out by the Scalable Game Design (SGD) team in the Department of Computer Science, University of Colorado Boulder, Boulder, USA as well as AgentSheets Inc. They have published several papers on their work, some of which are discussed in this section.

Basawapatna et al. [93] present a pedagogical approach to teaching entitled Zone of Proximal Flow. They combine Csikszentmihalyi's notion of Flow [94] with Vygotsky's theory of Zone of Proximal Development [95]. This shows how students learn in terms of skills vs challenges and they then apply this to their teaching methodology for SGD. They claim their principle idea of project-first, just-in-time ideas fits into this approach and encourages learning as students are given a challenge upfront that allows them to use skills as they learn them. They believe that this strategy will keep students engaged and this paper attempts to show its existence through showing that students attempt more difficult challenges. They hope that, with further study, this approach will be beneficial to lots of educational domains.

In their paper Repenning et al. [96] present a checklist for "computational thinking tools" as a combination of a computational thinking pattern [97] inventory, teacher training and authoring tool. They claim that to have an impact a CT tool used in K-12 should fulfil the following conditions:

- Have a low threshold (should be easy to learn)
- Have a high ceiling (should allow sophistication)
- Scaffolds Flow (progressing in sophistication should be as straight forward as possible)
- Enables transfer (can students apply what they learn to other scenarios)
- Supports equity (is it interesting & engaging)
- Systematic and sustainable (they need to be used)

They claim that, and give examples of how, Agentsheets meets these requirements.

Koh et al. [98] discuss a project in which students are taught game design by teachers taught by the SGD group with the concepts discussed by Repenning et al. [96] and Basawapatna et al [93]. in mind. They are focussed on whether the course is sustainable, in that, did students' progress in complexity and whether they transfer skills learned through game design to other areas? They claim that through their project CT skills learned and problem solving skills were extended and transferred. At the end of the 1-2-week module the student's creations were uploaded to the Scalable Game Design Arcade (SGDA). Evidence for this sustainability they give is:

- Probability to advance - over 80% of schools advanced further (more than they were trained to do) than the first project
- Number of different projects - on average more than three different projects were submitted by schools
- Advancing from game design to simulation design - 43% of schools moved onto science simulations and students

They believe this sustainability is a combination of student, teacher and school related factors and found in one survey that approximately two-thirds of students who took part would be interested in another game or simulation design course.

Nickerson et al. [99] present a way in which Agentsheets can be used to teach CT concepts using concrete and contextually appropriate examples. These concepts which they call Computational Thinking Patterns [97] (see Section 6) allow people to have a framework to describe phenomena that recur in multiple situations which allows students to use abstraction.

Worrell et al. [100] further discusses the SGD project and how it was used to teach students through a collaborative classroom design. The hope was that students would master and retain complex CT skills and apply them in assessments and that the collaborative nature would help them in this. They found that due to the context, where students may not be in the class for the whole semester, that the design worked very well and allowed late-comers to be integrated quite easily and retain a high level of information. After an initial introduction to some Computational Thinking Patterns as well as Agentsheets the students were divided into groups of two-four and given a specific task, such as make the game Frogger. Each student builds a specific part of the game/simulation and then teaches it to the rest of the group.

Basawapatna et al. [101] investigate how the SGD group integrate simulation and modelling into tasks and give different strategies for how this can be achieved. They show the link between simulation and CT, that in creating simulations users "start with a question, develop a model, express the model computationally, run the model, visualize the consequence of their thinking and possibly revise the model". They then given a "gentle slope" towards students creating their own simulations. This "slope" is made up of seven stages which are as follows:

- Animations – watching a movie or similar
- Interactive simulations – a simulation which the user can alter certain parameters
- Collective simulations – like above but with a social element
- Construction set simulations – construction kits used to solve domain-specific problems
- Pattern based authoring – begin to design the behaviours of simulations actors

- End user programming – using tools like AgentCubes or Scratch
- Traditional programming – using languages such as Java or C++

For each of the stages they give examples of how it can be done, tools to use and examples of how it benefits students as well as how it expands on the previous stage.

## 8.5 Lego Mindstorms

The Lego Mindstorms (https://www.lego.com/en-us/mindstorms) series of kits contain software and hardware to create customisable, programmable robots. They include an intelligent brick computer that controls the system, a set of modular sensors and motors, and Lego parts from the Technic line to create the mechanical systems. Mindstorms kits are sold both commercially and to be used as an educational tool, originally through a partnership between Lego and the MIT Media Laboratory.

Van Dyne & Braun [36] developed a CT course (CS0) at Montana Tech (University of Montana) which was used to assist students who were taking CS or Engineering as their major and struggling with maths skills. The course was primarily focussed on problem solving and critical skills and is now open to the general education curriculum. They taught the course using lectures and labs and during the labs they used Lego Mindstorms NXT robotics kits to illustrate the concepts taught in class. They originally used the graphical programming environment provided and then moved onto Java using LeJos, which is a Java library for the NXT robots. Topics taught using the robots included algorithms, data & variables, iteration and sorting. One example given in the paper is designing a "feral" robot which will back away if someone enters an area and then attack if it gets too close. This assignment incorporates looping and decision constructs as well as algorithm development. They found that their course increased student's analytical skills as tested by the Whimbey Analytical Skills Inventory [202] from pre-and post-tests.

Atmatzidou & Demetraidis [102] report on their use of robotics to teach CT and to see the impact of the activity on student's CT skills and to specifically see whether students of different age and gender develop them in the same way in the context of robotics. They emphasised algorithms, abstraction, decomposition, modularity and arithmetic operations whilst giving students questionnaires and getting them to write programs and run them using Lego Mindstorms robots. They found that CT skills significantly improved as the training proceeded and that students developed the same level of CT skills at the end independently of age (junior high -15 vs high vocational -18). The authors administered tests after four classes and after ten classes and found a significant improvement in results. They suggest that this shows CT skills need a considerable number of sessions to develop. They also found that boys and girls reach the same level of CT skill but that girls in general need longer time to reach the same skill level although this analysis was only carried out on one group - the junior high students.

Bers et al. [103,104] discuss their TangibleK course in which preschool-second-grade children are taught CT concepts through a robotics program. They used the CHERP (Creative Hybrid Environment for Robotics Programming) language alongside commercially available robotic construction kits. CHERP is a graphical language designed to provide young children with the opportunity to be introduced to computer programming. Topics taught include sequences, loops and branching programs through interactive activities including getting robots to dance the "Hokey-Pokey" and using light sensors to turn on its lights when it's dark. For this study the teachers used Lego Mindstorms. Students understanding of these different concepts were tested throughout the course and during the final project phase to see whether the students' scores on concepts changed with time and exposure. Students reached the target level of achievement but found the first half of the course easier than the second half which introduced more complex ideas; this could be due to course structure or perhaps the amount of time spent on each topic. Children also achieved higher score on their final project on specific topics (choosing and sequencing instructions). Their results showed, they felt, that the TangibleK robotics curriculum was engaging and developmentally appropriate for the children.

## 8.6 CS Unplugged

CS Unplugged (http://csunplugged.org/) is a collection of free learning activities that teach Computer Science through engaging games and puzzles that use cards, string, crayons and lots of running around. It was developed by the CS Education Research Group (http://cosc.canterbury.ac.nz/research/RG/CSE/) at the University of Canterbury, New Zealand (http://www.canterbury.ac.nz/), so that young students could interact with computer science, experiencing the kinds of questions and challenges that computer scientists experience, but without having to learn programming first.

Bell et al. [105] are the researches responsible for the CS Unplugged project and in this paper, they give an initial overview of the project as well as exploring why it has become popular and describe different ways it has been adapted which are listed below:

- Videos of different activities
- Making bracelets coded in binary
- Competitions
- Shows
- Adapting CS Unplugged activities to different themes such as WWII
- Outdoor activities
- Online activity

They also go on to discuss principles in place when designing the activities and discuss their future plans, some of which are referenced in this section and others in the paper.

Dwyer et al. [46] speak about the need to develop CT skills across the curriculum and their paper specifically focusses on an effort to reinforce physics phenomenon by designing games in Scratch based on physics concepts such as gravity, acceleration etc. The main research discussed, however, was done on a CS Unplugged activity called "Marching Orders" which involves one student giving another student a set of instructions on how to draw a shape. Through focus group interviews they found that fourth graders recognised the need for specific instructions but they struggled to produce them. They also recognised small errors could change outcomes.

Rodriguez et al. [106] assess the use of CS unplugged in middle school classrooms. They located a group of activities that were deemed age appropriate and through worksheets assessed their impact. Topics taught using the CS Unplugged material (adapted to suit 50 minute classes and to fit the age range) include Finite State Automata, Binary Numbers, Cryptology, Error Detection, Minimal Spanning Trees and Searching. Adding their own ideas and activities they then assessed how proficient students were at the topics based off the worksheets that were incorporated into the lesson. Proficiency varied greatly for each topic with binary numbers, for example, having above 80% proficient for all six questions, whereas searching had 53%, no comment is made on how that could be improved. They concluded that the CS Unplugged activities are good in and of themselves, but to be used effectively in a classroom they needed enhancement. Suggestions also include a priming activity in which students attempt to solve the problem in a naive fashion, that students need individual practice to fully grasp concepts, vocabulary on worksheets should be consistent with that used to teach the concept and that real-world concepts could help to engage students interest.

Pollock et. al [107] developed a course to train CS undergrads and trainee STEM educators how to integrate CS into their schools and curriculum and they primarily used Scratch as the tool to teach programming concepts. Undergraduate students would go out into schools to teach classes which used Scratch and CS Unplugged to teach CS and CT concepts such as variables, loops and conditionals. They also used HTML to design basic web pages. The students who participated (taught) in the course noted that CS Unplugged activities can require practising in their teaching and that scaffolding is often required without directly giving the answer.

Taub et al. [108] report on their study into the effect CS Unplugged activities had on 7th grade students view of CS. They also looked at the students perceived understanding of what CS is and their achievement of the task through both questionnaires and interviews. They were taught the Unplugged activity during classes or an after-school club for two hours per week. The topics covered included binary numbers, searching algorithms, error detection and sorting algorithms. The researches were interested to see whether student's viewed CS as problem solving, programming, how a computer works, fixing technical problems or as using the computer. They also looked at student's views of women in CS and what a career in CS is like. They also looked at student's intentions toward CS (whether they planned to study it) and their views as CS as fun/interesting. They found that students' intentions and attitudes regarding CS did not improve following participation in the activities and that their views of the nature of CS improved but not as much as the researchers had hoped. It was also found that their views towards CS as a career for women were good both before and after the activities. They also found that following the CS Unplugged activities student's view of whether CS is fun and interesting declined. This could be down to the ordering of the activities and that students did not fully understand the activities presented related to high school CS or a career in CS.

CS Unplugged was also found in many other course/curricula design as well as talked about as methods of teaching CT in several papers. CS Unplugged activities have also been adapted for various uses, papers that describe these or mention using CS Unplugged in their course/training include: [10, 11, 63, 109, 110, 111, 112, 113, 114, 115]

### 8.7 Other tools

As stated over 50 tools were found. Below is a list of more of these tools with the related papers that discuss their use, pros/cons and expand more on their design. Several of these papers are discussed in more details in other sections.

- Algo.Rhythm [65]
- Alice [7, 16, 29, 60, 73, 82, 109, 115, 116, 117, 118, 119]
- Ardublock [120]
- Arduino [13, 42, 65, 66]
- Bebras [121, 122]
- Binary toy [123]
- BingBee [124]
- Blockly [63, 71, 125, 126, 127, 128]
- Bunny Bright [123]
- CargoBot [129]
- CHERP [15, 24, 103, 104, 130]
- Code Bits [131]
- CTArcade [132]
- CTSiM [27, 133, 134, 135]
- CyberPLAYce [66]
- DigitMile [136]
- Dragon Architect [126]
- Drawing Machine Model [37]
- Entry [137]
- Escape Machine [138, 139]
- Game Maker [140]
- Greenfoot [97, 101, 113]
- HTML [8, 12, 107, 141, 142, 143, 144, 145, 146, 147, 148]
- Java [7, 9, 41, 42, 74, 101, 112, 118, 146, 148, 149]
- Lego WeDo [24]
- Lightbot [150]
- Lilypad Arduino [63, 120]

- littleBits [130]
- Logo [9, 17, 64, 138, 146, 151]
- Maple [152]
- MATLAB [18, 35, 153]
- Minecraft [42, 126, 143, 154, 155, 156]
- NetLogo [117, 125, 133, 135, 157, 158, 159]
- Pyonkee [129]
- Python [33, 40, 45, 47, 49, 50, 51, 58, 177, 125]
- RAPTOR [160]
- RuBot [124]
- Simulation Creation Toolkit [101, 161]
- STAGE [162]
- SUMO [134]
- The Incredible Machine [124]
- VPython [16, 39]

Other papers found during the search that give brief overviews of some of the tools discussed here, as well as others, may add to the scope for those interested: [13, 37, 39, 70, 117, 127, 143, 150, 163].

## 8.8 Conclusion

A huge number and range of tools have been developed to assist the teaching of CT. These range from music tools to programming languages to games. Although several of these are in the early stages of development, it is encouraging to see so many efforts to make CT fun and accessible to students of all ages, genders and abilities. The benefits for educators are many and include a variety of options of how to integrate CT into their classrooms. Whether in a computer lab, a regular classroom or outside, in a one-on-one session or with a class of 30+ there is a tool out there which will suit educators needs, and if there isn't then the evidence suggests that there might well be soon!

# 9 Research Question 5: How has CT been taught in educational institutions?

## 9.1 Introduction

The benefits of teaching CT have already been discussed, the tools available to do that as well as several ways in which it can be incorporated into already existing subjects. By far the most common way in which CT is being introduced is through Computer Science or Computing courses/classes/workshops. In this section, we will discuss a large selection and variety of these courses as well as how schools and higher education institutions have developed CT into standalone modules and courses. The hope is that those educators who have existing CS classes or who have time and the opportunity to accommodate CT independently will find ideas, encouragement and resources to help with this.

## 9.2 Teacher workshops

One way which has proved popular to facilitate the teaching of Computer Science and Computational Thinking has been through teacher workshops. In the following section, several of these will be discussed.

Blum & Cortina [164] present a pilot summer workshop (CS4HS) which was held in Carnegie Mellon University in 2006. This was one of the first attempts at teaching CT and making teachers aware of it. The workshop consisted of speakers on topics such as:

- The need for CT as a subject in primary and secondary schools

- The use of CS Unplugged in a classroom (included demonstrations)
- Career opportunities in CS
- How to teach principles of computation to students with no previous knowledge (included practical displays which included how to explore theoretical CS ideas using food)

On the second day panel discussions were held which focussed on the drop in interest in CS by students as well as more on job prospects and a session on computational biology which looked at DNA strings and matching algorithms. From results of pre-surveys and post-surveys, teachers found all the workshops and sessions were useful with the highest rated ones being the practical based sessions. They also found that teachers felt, after the workshop, that CS is to do with problem solving as well as developing CT skills for all aspects of life. This was a shift from before the workshop where most teachers said it was related to problem solving or programming.

Cortina et al. [116] present a work in progress on their project ACTIVATE (Advancing Computing and Technology Interest and innoVAtion through Teacher Education). They invite teachers to attend summer workshops where they are shown how to incorporate computing concepts into existing STEM courses. These were week-long workshops for STEM teachers where teachers received four days of instruction in programming and computational concepts. On the final day, teachers designed an application in either Alice, Python or Java to illustrate these concepts in curricular topics they currently teach. The workshops were all well received and follow-up surveys and interviews indicated that over 90% of teachers planned to use the workshop materials in either new or existing courses.

Curzon et al. [165] present their approach for workshops where they introduce CT and programming concepts to teachers through "unplugged" methods in a series of four workshops. A brief description of each of these workshops are:

1) CT - this included four activities split into two halves. The first half consists of two activities; the first uses the context of helping someone with locked-in syndrome. Participants must communicate by just blinking. The participant blinks when they encounter the letter of their choice, which allows them to spell out words. The idea is then to improve this method to be quicker and more efficient. The second activity is playing a game of 20 questions, which teaches divide and conquer algorithms. The second half is composed of an algorithmic based magic trick and a final activity shows how this can be applied to binary encoding on a punch card.
2) Algorithmic thinking - the first activity is playing a game of X's and O's against a piece of paper. The paper has instructions on it of how the paper should "play" and the paper usually wins the game. The second is a magic trick while the third is about how there is more than one solution to lots of problems using solitaire style puzzles.
3) Unplugged programming – the first activity is programming a robot face based on noises the audience makes - the robot is a person. The second activity is an activity on variables and assignment using boxes and coloured paper. The third activity is a simple if-based program where audience members represent a command or expression.
4) The Human Side – this workshop emphasises the need to understand people when using CT. To do this they use a card trick, teach medical device design and show a video of a game designed to teach user testing.

Morreale et al. [112, 149, 118] discuss a series of summer workshops in which the aim is to change the perceptions of educators and students toward CS. They include a list of topics taught as well as results from surveys. In one such day-long workshop topics covered included Computational Concepts with Alice, CS Unplugged: Computing without a Computer and Easy Java Simulations. The workshop was well received and deemed successful in that they changed teacher's perspectives of CS with a 27% increase of how many teachers would recommend students pursue a career in CS or IT and a 30% increase in recommending students attend one of these workshops. Another, week-long, workshop had teachers attend with up to four students from their schools. The tools used included Easy JavaSim and POV-Ray and the activities included Easy Java Simulations, Bouncing Ball and Harmonic Oscillation; these were based off Shodor [166]. They found that the results were very encouraging with students that attended realising there was more to CS than they thought and that it can be fun and applicable. Teachers perceptions were changed and teachers responded that it should be included in the curriculum as well as seeing a 50% increase in the number of teachers that would recommend CS or IT as a career to their students.

Morreale et al. [112] look at the impact of CT workshops on high school teachers. Teachers took part in two workshops, one in the summer and the second in the autumn and filled out pre-surveys and post-surveys. They first found that most educators correctly identified what CT is and this percentage increased from 75% to 89.65% after the first workshop. However, less chose the correct answer for why it is important (which they stated was "A way of thinking and problem solving") but again this was increased by over 20% after the workshop. Approximately 10 teachers attended both workshops and they were asked if they had used any of the given sessions in their classroom. Four of the eight workshops given were identified as immediately useful with the most used being a presentation on CS careers, others included Algorithmic thinking and CS Unplugged. They state that the use of these is positive but more work is required to make all the sessions accessible to teachers.

Vieira & Magana [167] discuss the outcome of a three-day workshop which was aimed at introducing teachers to CT practices, concepts and principles. It was based on the CS Principles which is part of a U.S initiative aimed at developing or supporting existing Advanced Placement (AP) courses. The guiding research question was "How teachers implement the backwards design process embodying elements of CS Principles in the context of their classrooms?". They were hoping participants would be able to identify different approaches to implementing CS principles and use Backwards Design to design learning activities. They also taught participants Scratch. Sessions included initially discussing CT and what it is, as well as what CS is and the CS Principles. There are eight parts to the CS Principles frameworks, namely:
- Creativity
- Abstraction
- Data
- Algorithms
- Programming
- Internet
- Impact

The steps involved in Backward Design were then introduced as well as a discussion on learning environments such as Scratch and WISE. Participants then designed and presented a lesson plan on at least one of the CS Principle's learning objectives. They found the CS Principle's framework useful for designing the activities but that participants mainly focussed on one of the eight available, "computing as a creative activity". They also believe that through the workshop CT concepts became "clearer to all participants".

Pokorny & White [111] give an overview of their CS4HS workshop. They give an overview of the workshop which included presentations on the Computer Science Teachers Association K-12 standards, recruiting women into computing and careers in technology. Hands-on sessions were also given including sessions in CS Unplugged, computer hardware and Scratch programming. They also included sessions for participants to create "action plans" to incorporate CT into their school's curriculum. During discussion sessions, it was found that participants had little to no understanding of CS concepts but by the wrap-up session teachers responded that they had a better appreciation for CS and CT and of how it could be incorporated into their subjects and schools. They found that there was a high level of satisfaction with the workshop and that most would attend future events as well as recommending it to colleagues. Five months after the workshop a second survey was distributed to see the impact it had had on teachers. Several had incorporated Scratch and Excel programming as well as discussing CS careers with their students. Others spoke of difficulties including lack of support from IT departments (for example when wanting to use Scratch) as well as a lack of time to refine lesson plans and develop competency in the material.

Falkner et al. [168] discuss their development of a MOOC course to support teachers in implementing a newly developed computing curriculum in Australia. They give an overview of their course in which they focus on teaching teachers about computation concepts whilst being tool-independent. Units included:

- Data - Representation/Patterns & Play
- Digital Systems
- Information Systems
- Algorithms & Programming
- Visual Programming.

Each unit contained two fully worked examples of how learning objectives in the K-6 curriculum could be met. Analysis of data from educators who have used their course found that it could assist teachers in developing confidence and improve their understanding of CT and digital technologies. They also found that giving everyday examples and cross-curricula connections of CT concepts and programming statements helped teachers feel more comfortable with the new curriculum.

Yadav et al [169, 170] discuss introducing CT to teachers currently in training. The authors ran a one-week module in which students were taught CT concepts such as problem identification and decomposition, algorithms and debugging. The teachers in question had no prior CS background and so were taught these concepts using day-to-day examples, some of which are given in the paper. The second class illustrated how CT could be applied in an educational setting, examples included role-playing and simulation (for example with science). Participants of both a control group and the experimental group filled out a quiz. Teachers who received the module thought of CT as a problem-solving approach whereas the control group tended to agree with the idea that CT involves working with computers. The participant's attitudes towards computing didn't change significantly however the authors state this could be down to them realising their lack of understanding and so having no more confidence in it. In [169] the authors present further work on a CT module they designed to be taught to all elementary and secondary education majors. The course was introduced during a "Learning and Motivation" module and was designed to go alongside the already existing topics such as: formal & informal assessment and learning styles. They introduced students to the definition of CT and five concepts and then focussed on CT in day-to-day life. They also gave a demonstration of teaching algorithms through kinaesthetic activities and showed a recursion example using the Towers of Hanoi. Surveys suggest the course was effective in increasing student's awareness of CT and how it can be better integrated into their future teaching by promoting problem solving. They also found that the student's attitude towards it became more favourable.

### 9.2.1 Conclusion

From these various papers and studies there are several ways in which teacher's enthusiasm for, knowledge of and ability to teach CT and CS in their classrooms can be improved/increased. Most popular seem to be day-long workshops and workshops that are heavily practical in nature. The ideas, tools and lessons that are given during these workshops seem to give teachers a greater understanding of what CT is and how it can be useful for their students whilst also giving them very practical ways to implement this in a variety of contexts. Wide-reaching initiatives such as Google's CS4HS can help teachers that otherwise lack the skills and knowledge to teach CT topics.

Interestingly it seems that one significant barrier to CS and CT in education is teachers and educator's misconceptions about what these are. One advantage of having teachers attend these training days and workshop is that these misconceptions and misunderstandings can be corrected, which is successfully done in most of the described studies. It can also be seen from these papers that teacher's willingness and interest in teaching CS/CT is vital in its implementation in both primary, secondary and tertiary education.

## 9.3 Early Education

Although not searched for explicitly, two studies on introducing CT into early education classes. These are discussed below.

Lu & Fletcher [4] discuss how CT and CS should be taught, if it is to be foundational learning goal on par with reading, writing and arithmetic. They suggest that CT should be introduced first by establishing vocabulary and symbols that can be used to describe computation and abstraction. The main point of their paper is to suggest that programming, although an element of CS and CT, shouldn't be the first thing taught, but should be a "entrance into higher CS". To this end, they suggest a computational thinking language (CTL) and given examples of how it can be taught including an introduction to multiplication and charting information. They also give some notation and ideas for advancing and integrating it, again giving examples of both including finding square roots and interdisciplinary projects.

Portelance & Bers [171] report on a novel technique for assessing the learning of CT in early childhood classrooms. Students in second grade classrooms were taught basic CT concepts using ScratchJr and then conducted video interviews in pairs using their iPad cameras. The children would develop Scratch projects and then three interviews were conducted in pairs with one of the students filming whilst the other explained their project to them. Questions designed by the researchers were displayed prominently and were as follows: Tell me about your project. How did you make your project? What would you do if you had more time? Question of your choice. Giving examples of interview excerpts, they show how children learnt concepts such as reuse, sequencing and space cost. They also give some recommendations of how to implement this kind of interview activity: 1) Do a practice run, 2) Demonstrate good camera work 3) Demonstrate good presenting. They conclude that students can show a broad range of CT concepts and about their projects. They state that the interviews seem to "capture rich data aimed at contextualizing projects and understanding children's thinking".

## 9.4 Primary School

Although not as common as secondary school studies, several papers were found in which CT is discussed in a primary school context. This section will discuss a few these.

Falloon [129] presents findings from a study of nine and ten -year-old students using two apps to undertake coding tasks. The idea of the study was to learn whether these apps were useful for developing CT. The apps in question were CargoBot (https://twolivesleft.com/CargoBot/) and Pyonkee (http://softumeya.com/pyonkee/en/) and they were used on student's iPads. The author gives an overview of the research context as well as an overview of the apps. Using a coding system based on Brennan and Resnick's framework [172] Studiocode video analysis software was used. An in-depth examination of this is given and results are provided which showed that both apps are useful but for developing different dimensions of CT. The author states that CargoBot might be more efficient in developing a more technical understanding of computational concepts. They also noted that CargoBot presented less distracting options, that Pyonkee's design appeared to stimulate higher levels of collaboration and that more time was spent testing and debugging in CargoBot. They also noted was that with both apps there was little trial and error which suggested problem solving and debugging was a deliberate and reflective process.

Miller et al. [151] presents a study into the problem-solving and spatial relations of fifth and sixth grade students who had one academic year of experience using Logo. Two programs, The Factory [203] and Teasers by Tobbs [204] were used to assess problem-solving and spatial relations ability was measure by subtests of the CTMM (California Test of Mental Maturity: Level 1) and PMA (Primary Mental Abilities Grades 2-4). The two programs are described as well as the demographics of the participants. Results showed that there was significant difference between the treatment group and control groups in both problem-solving measures and one of the three spatial tests. However, no pre-tests were given so it cannot be assumed that one group learned less or more.

Bell at al. [173] designed a course in CS/CT for primary school. They found, from teacher feedback, that teachers could deliver the material in an engaging manner and often better than expected. Teachers also found many opportunities for cross-curricula teaching with subjects such as Maths, PE, Creative Writing and Art. Teachers also commented on teamwork, cooperation and communication and noted that computing was helping with those skills.

Chiprianov & Gallon [174] present an overview of a project to integrate CT into French elementary education. They give an overview of the present state of CT Education in France and talk about child cognitive and affective development. They then discuss the theoretical framework of the curriculum. The new French curriculum included programming and is in line with the CSTA standard but does not explicitly include CS. They then give an overview of tools and activities they plan to use to teach CT, which includes games (code.org) and robots. Furthermore, they give an overview of putting it into practice, namely the development of 10 lessons and give some strategies of teaching which was usually as follows:

- Whole-class discussion and demonstrations, followed by
- Collaborative or individual learning, and finally
- Reflecting on the solutions and discussing these

Lastly, they give an overview of some lessons they learned which included their belief that teachers are important for both a positive attitude as well as to adapt the materials to suit their classrooms. They also found that the children acquired and learned skills, such as space location, during the given tasks.

Jovanov et al. [136] present an overview of the newly introduced Macedonian curriculum entitled "Working with computers and basics of programming" or "Computing" for short. They give an overview of the state of computing education in Macedonia prior to the curriculum and then give an overview of the new introductory subject for pupils aged eight. They give an overview of the content that includes seven units to be taught in two classes per week:

- First steps of using the computer
- Computer graphics
- Text processing
- Online living
- Notion of algorithms and programs
- Computational thinking through a game
- Creation of simple programs

They go into detail on the final three which are focussed on CT. Students are taught the notion of programming and learn through a game which was specially designed. They give an overview of this game called DigitMile, which is designed to be used alongside the curriculum. Finally, students use ScratchJr to develop simple programs. They also detail teacher training and state that prior to training teachers were apprehensive about the changes but were more confident and understood the need for the topic after it.

Mensing et al. [127] give an overview of how the Common Core Standards introduced in the US have a lot of relationships between them and coding concepts and how they developed their "Coding is Common to the Core" initiative. Examples include the following:

- Mathematics - Logic and Sequential Directions (Blockly Maze) e.g. discretisation movements of sprites/objects
- English Language Arts - Blogging and Idiom

They then give an overview of some tools used to teach coding such as Scratch, Codecademy and Blockly. They also give some classroom integration examples.

Calderon & Crick [175] present sessions in which they aim to teach both Human-Computer Interaction (HCI) and CT to primary school students. They first cover negative transfer and the principle of affordance. The session uses a "doll house" designed in a "self-contradictory" manner. Students are given clues as to how to navigate the house and their goal is to find a pebble. An example of affordance and negative transfer is the fact that past experiences affect the children learning new tasks through, for example, the doors. The first instruction is to enter the house through a door that has a door knob on it; however, instead of being a door that opens through pulling (which the doorknob would suggest) it opens when pushed. The next door also has a door knob but this one needs to be pulled. The second session teaches Design for recognition and involves students being given cooking ingredients to transform into symbols. One group is given lists of portions, instructions and ingredients and they must transform these into symbols so another group can put a recipe together. Iterative design can be added to both sessions by letting students have second and third attempts at the sessions.

Sabitzer et al. [114] discuss how informatics topics are "hidden" in the Austrian primary school curriculum. They say that most teachers are unaware they are already teaching informatics and give a sample project on CT and they report on different initiatives that aim to introduce informatics to primary schools. One such project is "Informatics - A Child's Play" and it is partly based on CS Unplugged and Informatik erLeben [176]. The project includes the development of teaching units and materials and the implementation of these in schools. Initiatives as part of the project include teacher training as well as inviting children and adults to the university's lab. They then discuss the IMST project "Aspects of CT in Primary Education" which focused on algorithms and problem solving in primary education.

### 9.4.1 Conclusion

It can be seen from these papers and others discussed in other sections (see Section 8) that there are many methods and opportunities to teach CT to primary school students. Visual programming languages such as Scratch/ScratchJr allow young children to be introduced to programming in a user-friendly environment. Other concepts can be taught through practical, hands-on activities such as CS Unplugged. It is also important to show teachers how these can fit into other subjects, or how they may currently be teaching it already through other topics.

## 9.5 Secondary School

Secondary school aged students appear to be the focus of much research into CT and CS in schools. Across the world there many curriculums focussing less on programming and more on CT and concepts have been implemented. Most of these have been well received and we will now discuss several of these.

Bargury et al. [148, 145] present their curriculum for Israeli middle school in which they hope to expose students to the fundamentals of CS and CT. At the time, CS was included in high schools as well as a software engineering course. The aim of this new middle school course was not to make students programmers, but to teach them logical and algorithmic thinking whilst exposing them to programming. The course contains four modules taught over three years (180 hours, two per week), each of these modules is summarised as follows:

- Module 1 - Exposes students to fundamentals of CT and programming such as loops, execution variables and event handling. Scratch is used to teach this
- Module 2 - Scientific research using spreadsheets (required for maths and physics so cross-curricula)
- Module 3 - Elective: Introduction to Robotics, Basic Internet Programming (HTML5 and Javascript)
- Module 4 - Programming project (includes a proposal, modelling a problem, designing and implementing a solution)

Sentance et al. [177] discuss the challenges involved in the introduction of CS in UK secondary schools as well as the progress made and support provided for teachers. They give an overview of CS education in the UK and how it has gone through changes, beginning from CS-related topics to a shift towards ICT which had a focus on the use of software. Since then the shift has been in the reverse, to a more CS-focussed course. One challenge is the recruitment of teachers and they discuss various other countries standards, including Israel where teachers are required to have a CS degree. They then discuss teacher development in general, including mentoring, classes etc. Specific challenges in the UK include upskilling existing teachers, training new teachers and curriculum and resources. They go in to detail in the first of these where a survey was distributed to teachers who wished to attend professional development courses/training in CS. Areas where teachers sought support were "guidance on ways of teaching Computing" and the need for resources and tools. They also found that one-day workshops and working with an experienced teacher were the most helpful kind of development. Time was the most significant area in regards to teachers and school's willingness to participate in these. They present some examples of these training courses including Python School, Digital Schoolhouse and Computing at School Master Teachers. They then discuss two aspects for future work which are accreditation and action research which includes workshops, a network of excellence and online forums.

Brancaccio et al. [152] discuss the Problem Posing and Solving project which aims to enhance teaching and learning in Informatics and Mathematics in Italian High schools by using an e-learning platform. They give an overview of how this was developed (using Moodle) and then give some activities of how they have implemented some scenarios. They state that the methodology is closely connected with CT and one example they give is how CS teachers have adopted a "Living Lab" approach where they focus on a problem that is easily understood and perceived which they can then use as a reference framework and use tools such as Python to implement it. They also give an overview of training being distributed over this system for teachers.

Curzon [110] discusses the "Computer Science for Fun" (cs4fn) project in the United Kingdom in which the aim was to inspire students and teachers to learn and teach CS as well as give them the tools to do so. It consists of a magazine, website, live shows and support for teachers. The aim was originally for secondary school students (14+) but has since broadened to include younger students. They have found enthusiasm and interest for the project have increased and they say that teachers and students have given positive feedback overall. They move on to discuss CT and how they have introduced it to their course, mainly through "Unplugged" activities. Some examples they give are CS magic shows and Locked-in Syndrome.

Carvalho et al. [178] present a paper on difficulties in the deployment of CT in education, specifically in their context of Brazilian Basic Education. They give an overview of CT as well as some examples of other courses and initiatives to promote CT in education. They then give some challenges the Brazilian education system would have incorporating CT. These are:

- Lack of infrastructure i.e. lack of computers
- Rearrangement of the curriculum i.e. new subject or incorporated into others? Due to the structure of educational institutions in Brazil either of these is possible as it is quite a flexible system
- Qualification and backgrounds of instructors i.e. do they need a computing degree?
- Strategies to disseminate CT i.e. workshops, textbooks etc.

Riberiro et al. [179] also discuss some challenges and possibilities when it comes to including CT into Brazilian schools. They agree with Carvalho et al. that curricular changes would be needed and the training of CS educators is vital. They also cite Government Policies as a potential issue and they remark that it is important the government build workgroups to lead the inclusion of CT into education. They also say that people, and traditions in learning, mean that people might be reluctant to take up CT as they might be hard to convince that it is a skill they don't already possess.

Li et al. [180] present three activities designed to teach CT to high school students. The three activities are presented and described, along with the concepts they hope to teach. These concepts are:

- Looking out for the thief - Binary search
- Weighing fruits - quick sort
- Drawing geometric figures - using Scratch, iteration

Prior to these three activities, an introduction to CT class was given to the students. Post-test results showed that students were capable of decomposition to solve the problems as well as the first two activities being interesting and helpful to learning, with the third only being helpful whilst being boring.

Caspersen & Nowack [181] present a new and generic approach to Computing in Danish High Schools. It is based on a framework that they constructed from ideas based on ideas related to CT. They give an overview of computing in Danish High Schools before presenting two thesis' of which it is based and then give an overview of the modules (called "Knowledge Areas").

The thesis' are that:

- "Through computing people can create and handle thoughts, processes, products and services that create new, effective and border-crossing opportunities - impossible without digital technology."

- "There exists a common and shared foundational set of computational concepts, principles and practices, which can be applied purposefully with science, arts and humanities, and health and life sciences."

The Knowledge Areas are as follows:

- Importance and Impact - relevant and significant areas computing/IT is used
- Application Architecture - IT architecture of systems (presentation, logic, data)
- Digitisation - data representation and manipulation
- Programming and Programmability - intro to programming
- Abstraction and Modelling - modelling data
- Interaction Design - describe and analyse interface design, implement it
- Innovation - product and process perspective

They then give an overview of the pedagogical approaches that they have based the framework on, namely:

- Application-orientated - top-down rather than bottom-up, their aim is not to teach specific competencies (e.g. programming), but developing interest, critical thinking and CT
- From consumer to producer - students begin as users of an artefact by using/studying it, then modify it and potentially create from scratch (use-modify-create)
- Worked examples - problem statement and a procedure for solving the problem

The DISSECT project is discussed in several sections of this paper (see Section 8 and Section 9.2). In their paper on the project, Burgett et al. [59] give an analysis of the pedagogical techniques used to integrate CT. DISSECT pairs graduate "fellows" with teachers in the aims of developing lesson plans that incorporate CT and these modules are designed to show how CT already exists in students' lives. These modules are then varied and made appropriate for the different contexts schools and teachers find themselves in and creativity is encouraged. They then present results on two terms worth of these courses being run and have found that in general interest, motivation and knowledge of CS was increased in the students. They note that further studies are required to see if students learned/mastered the concepts. In their paper Folk et al. [182] give a summary of different modules developed for middle and high school classes through the DISSECT project. Module titles include Fingerprint Generation Algorithm, Weather Tool Algorithm, Binary Search and Diagramming. Results from these modules are given which in general show a statistical increase in CT knowledge and qualitative data shows that they were successful and that students understood concepts and had fun learning them.

Grover et al. [74] expand on their FACT curriculum (see Section 8.2) and in this paper, give a more thorough overview of the background and research behind the curriculum. They also give an overview of the units that FACT is composed of. These are (paraphrased):

- Computing is everywhere! /what is CS?
- What are algorithms and programs?
- Iterative/repetitive flow of control in programs
- Representation of information
- Boolean logic and advanced loops
- Selective flow of a control in a program
- Final project - chosen by students, individual or in pairs

They also give the learning outcomes of each unit and give an overview of the assessment tools used, which include Scratch assessments and quizzes as well as some sources of questions used. Analysis from these assessments as well as interviews and other qualitative data show that FACT helped learners understand algorithmic flow. Two separate studies, one a face-to-face version of the curriculum and the other an online version showed that the online version worked as well if not better than the face-to-face. They also found that the seven weeks using Scratch allowed students to be able to interpret text-based programs quite successfully. They believe this is due to the algorithmic concepts taught to students allow them to take that skill and adapt it to other tools, which is one of the aims of FACT.

Hazzan et al. [183] present a model program for high school CS education based on the Israeli high school curriculum. The model consists of four components and the relationships between them. The four key components are:

- A well-defined curriculum (including written course text books and teaching guides)
- A requirement of a mandatory formal CS teaching license
- Teacher preparation programs (including at least a Bachelor's degree in CS and a CS teaching certificate study program)
- Research in CS education

These are expanded on in more detail in the paper as well as the connections between them.

Pasternak & Vahrenhold [141, 142] give details about their proposed approach for implementing CS curricula in secondary schools which they call "braided teaching". The idea is to have the subject organised by either structure or content. They then give an example of this method of teaching using a strand that contains items related to semi-structured data such as XML. They also talk about how programming can be taught as a strand, the idea being that it, along with all other topics, are mixed with one another and no topic is more or less important than another.

Towhidnejad et al. [154, 184] present some activities they have developed to introduce sixth to 12th grade students to CT concepts and "entice" them into recognising that they can understand computing and engineering topics. One way in which they do this is through introducing CT into topics which are not related to computing such as Chemistry and Physics. In the former they discussed the Fukishima reactor failure and covered Fault Tree Analysis. In the latter paper [184], they discuss music recording with concepts such as sampling rate and memory devices. Another way which they have taught CT is through games and they describe how they used Minecraft to teach topics such as Finite State Machines and Shift Registers. They also give a description of other games that do not require a computer, these games teach topics such as searching and sorting and bitmaps.

Some comments they make based on their four years of introducing CT are as follows:

- Research shows that the earlier we can introduce students to computing the sooner we can get them attracted to the field
- They believe it has several side effects such as:
    - building confidence in dealing with complexity
    - dealing with open-ended problems

They acknowledge the problems with introducing CT into schools and so they describe the above examples as "stealth teaching", where the technical concepts are hidden to begin with and the students interest is caught by the topic first.

Rode et al. [120] present their educational framework in which they aim to move from CT to Computational making. They define computational making as a combination of making, which is the act of creating tangible artefacts and CT which they use several previously given definitions. Their method was to use e-textiles and LilyPad Arduinos to introduce children to CT. After giving an introduction of the LilyPad to the students, all of whom had some sort of computing knowledge, a Bunny Electronics kit was used to teach them about the circuits. In the second activity, children made interactive stuffed monsters using the LilyPad. They give more detail of the above sessions including how they used Ardublock, a block code interface similar to Scratch. They then give examples of how the different processes in making the monster connected and taught CT, for example Data Abstraction when the students had the opportunity to explore the LilyPad and alter the LED blink frequency. They conclude that five key factors were involved in the activities, namely: Aesthetics, creativity, constructing, visualising multiple representations and understanding materials. They also recommend a gamification approach to teaching CT.

Mooney et al. [185] present the PACT initiative which is a partnership between Maynooth University and secondary schools in Ireland. They used the name PACT as an acronym for Programming and Algorithms together teach CT (Programming ^ Algorithms = Computational Thinking). After giving an overview of the state of computing and computer science education in Ireland they then give a description of their pilot study. This included school teachers being given material on five different units to teach to Transition Year students. These were Programming 1, Programming 2, Algorithms, Graphics and Recursion. Teachers were given a two-day workshop during which they were introduced to the resources and carrying out programming exercises and practical exercises for all areas of the course. Feedback was then given in the form of a survey from eight teachers and 61 students. They found that students found the material challenging and that both teachers and students felt a less theoretical approach might be helpful when introducing key concepts.

Roscoe et al. [186] report on a series of workshops that they have developed to teach CS to students in a variety of settings including multi-day camps and three-hour school workshops. One workshop involved using Printcraft to teach computational and engineering concepts without using the complicated CAD software. Students are given the task of constructing a building or buildings (school, houses etc.) in a Minecraft type environment and this is then made using a 3D printer. Another workshop used App Inventor to teach students concepts such as loops, variables and conditionals. They comment on how graphical programming allows algorithmic skills essential to CT taught in a way that is not as intimidating to students as a formal language (Java, Python etc.). They also used robotics, specifically Arduino and Lego Mindstorms to teach students about software, hardware and their interaction as well as understanding circuit boards. These workshops could be adapted to include content based on CT concepts such as conditionals, abstraction, algorithmic thinking and problem solving skills.

Huang et al. [187] present a teaching method for programming courses in which they try to emphasise the importance of CT and problem solving and that a programming language is a tool for this. They give examples of how this can be implemented by suggesting that the focus be on problem solving first and then suggest an object-first approach, which involves students learning about objects before learning a programming language.

### 9.5.1 Conclusion

As can be seen from these studies, researchers have gone about introducing CT and CS into secondary schools in a variety of ways. Some implement full curriculums whereas others focus on optional courses and modules. Some common themes include:

- difficulties such as lack of skilled teachers (or lack of training for existing teachers)
- lack of equipment, especially in poorer countries
- a wealth of recourses and tools are available to assist teachers who wish to teach these concepts to students

It is also clear that different concepts and tools are more appropriate for different age groups, with certain tools suiting younger students (Scratch for example) and some concepts being better taught later in education (higher-level programming). However, there is a wide range of concepts and topics, which can be used to improve both students CT skills and their enthusiasm towards and understanding of CS.

Encouraging inroads are being made into making CS and CT more accessible, widespread and teachable in secondary schools.

## 9.6  College/University

Astrachan [188] gives an overview of their course which is delivered at Duke University, North Carolina, US. Their course is not a conventional CS course in that it does not involve any practical, hands-on work but is based on active-lectures and discussions. They call the approach "pander-to-ponder" by which they mean that students are first "pandered too" by not only making the course accessible and using popular media but also first showing students how CS really affects the world. They then encourage students to "ponder" the topics discussed which include Encryption & the 5th Amendment of the US Constitution and Network Neutrality & TCP/IP. An example from the Encryption idea is that they want students to understand the impact of computation and why the techniques (such as modular arithmetic) are used rather than making them do the mathematical calculations.

Ater-Kranov et al. [189] present the development of learning modules, initial findings and challenges of the NW-DCSD (Northwest Distributed Computer Science Department) project which aims to offer an "innovative and inclusive vision of computing in the 21st century". They give a definition of CT and then give some examples of CS and Non-CS modules which they hope to promote CT. These modules are:

- Internet Connectivity (CS)
- Robot Defence (CS)
- Airport Traffic Simulation (CS)
- Algorithmic Art (Non-CS)
- Animation of Proteins/Cellular Components (Non-CS)
- Algorithmic Composition (Non-CS)

Using surveys and knowledge probes, they give some findings on the six modules named above. Their findings found that there was some commonly agreed upon skills that are viewed as important to CT (e.g. Using critical thinking) but that there is no universally accepted definition. They also found that most CS students agreed with statements such as "CS requires a lot of creative thought" but more importantly that after the courses, more non-CS students strongly agreed with the same statements. Overall, the students found the modules interesting and that they learned a lot from the assignments.

Czerkawski [190] discusses how CT can be integrated into a virtual higher education curriculum and gives some thoughts on pedagogical considerations. They give information of how CT tasks can be implemented in a virtual classroom by first giving an overview of the ISTE (2011) definitions of CT concepts and how they can be integrated into virtual learning. These include:

- Data collection -> Create an online survey
- Data analysis -> Use excel (or similar) to analyse data
- Data representation -> Create social maps to summarise findings
- Problem decomposition -> Create the first blueprint of instruction design plan
- Algorithms & Procedures -> Work as a group to create a teacher guide and FAQ

In these activities, students are not engaged in programming but are introduced to the vocabulary and mental tasks of computation. The author notes that these could be taught in a physical classroom but that the online nature allows for use of social collaboration and communication software. In this way, it might be that students get used to using technologies and then the transition to more "traditional CT teaching i.e. programming" might be easier.

Dierbach et al. [191] present their experiences of developing a model for integrating CT into the general education curriculum for undergraduates at Towson University. After initially finding there was interest they ran workshops to develop a common set of course objectives and chose four discipline-specific courses. The courses were the following: CT in the humanities (Dept. of English), CT: Developing Life Skills for Weight Management (Dept. of Kinesiology), CT: Creative Work with Audio and Video (Dept. of Music) and Revolutionary Networks (Dept. of Sociology). The common course objectives were:

- To develop and/or evaluate computational models and apply these models to a problem or domain
- To create and/or apply algorithms, and determine the appropriateness of algorithmic techniques for a given problem
- To distinguish between problems that can and cannot be solved computationally, either theoretically or practically
- To evaluate a given algorithm or model, and specify appropriate criteria used

- To apply and reflect on the application of established computational thinking methods and models to tasks, problems and investigations in specific domains or disciplines

They also give an overview of each of the modules specific learning goals. Finally, they give an overview of an initial assessment of this course. They found that faculty members showed interest in the project and that most students found the material interesting and relevant and there was positive feedback from administrators.

Soh et al. [192] give an overview of their project which is entitled "Renaissance Computing". In it they propose a re-think of how CS is taught in universities, that it is linked to other domains and so should be targeting and including both CS majors and non-CS majors. They give an overview of how their thinking can benefit both CS and non-CS majors and why they are pro collaborative learning. They then give an outline of the framework they have developed which includes separate CS1 courses but a unified CS2 course and a final capstone course in which other disciplines will work on group projects with CS majors and minors. They then give an overview of the planned roll-out of the courses.

Hurson & Sedigh [193] give an overview of the changes they hope to make to transform the teaching of computing to engineering students. They feel that most of the beneficial problem solving techniques and CT are lost through traditional programming courses and so engineering students lose potentially useful skills. The course broadly involves a common lecture introducing the concept with domain-specific problems then being given in laboratory sessions. The topics include the following, all of which they consider to be foundational to CT in engineering:

- Algorithms
- Validation
- Verification and correctness
- Maintainability
- Reusability
- Fault tolerance

They also give some examples of the domain-specific problems, which include Aerospace Engineering Project on Rocket Trajectories and a Civil Engineering Project on Load Bearing. In conclusion, they give an overview of how they will gauge the effectiveness of the course as well how it will be assessed.

Kafura et al. [125] give an overview of their CT course. The course is divided into four parts, namely:

- Computational Modelling (4 weeks)
- Fundamentals of Algorithms (2 weeks)
- Data-intensive Inquiry (7 weeks)
- Social Impacts (2 weeks)

They used a variety of tools including NetLogo, Python and Blockly. The course aimed to be taught with collaboration, context-based learning and transference at its heart. Some of this was achieved by an open-source e-textbook platform named "Rhinestone" which they developed and was based on Runestone [194]. It allows students to work collaboratively like in Google Docs or similar, and supports Blockly and stores all changes made by students. Some conclusions they make include that the course engaged and motivated students and that multidisciplinary cohorts were well received and fostered learning across disciplines which was one of the goals of the course. They also found that students had a deeper appreciation for computing in their disciplines as well.

Shell et al. [155] present their Computational Creative Exercises (CCE) which they designed to be used in a CS1 course for engineering students. They begin by showing the link between computational and creative thinking skills in these CCEs such as: Algorithmic Thinking -> Challenging accepted solutions and procedures. This paper looks at a quasi-experimental trial in which the impact of the CCEs on learning, engagement, self-efficacy and other metrics are measured. Ninety students took part in the study by completing pre-surveys and post-surveys with a control class of 65 students. They give specific examples of how they measured the metrics. They found that CCEs positively impact student learning of CT and CS knowledge and skills. The students who took part in the CCEs also had a significantly higher self-efficacy for applying their CS knowledge in their field and those students also had a significantly higher study time and lower lack of regulation (measures difficulties with study). They conclude that including CCEs in CS courses could improve CT and learning of CS concepts and skills.

In [195 and 196] Miller et al. also discuss CCE's and their design. Their exercises have four key components:

- Objectives - this lists the computational and creative thinking objectives for the exercise
- Tasks - the main content, group activity, designed to involve all and provoke discussions
- CS Light Bulbs - self-contained explanations which link the tasks and CS topics.
- Questions - open-ended to engage students to analyse and reflect

They describe four CCE's they have designed under those four components in detail. The CCE's are called Everyday Object, Storytelling, Cipher and Exploring. After analysing data from course grades, CT test etc. they conclude that the exercises appeared to benefit student's achievement and learning with both long-term retention and grades improving. However, it was noted that students didn't always recognise the benefits and there was a need to more specifically explain how creative thinking helps CS and how it relates to CS topics.

Cortina [160] talks about the design of an introductory CS course for non-CS majors. The course focuses on the process of computation and does not include programming; instead, students focus on algorithms and CT. They list the course topics, which include the following: "History of computation", "Expressing computation using algorithms", "Organising data", "Computational tricks", "Applications and the Future of computation". As a lot of algorithms are presented students need to be able to see them work to understand how they work and to do this the authors used Raptor which is a flowchart building tool. They then give a detailed overview of the students that participated in the course over the first three semesters. Some findings from student feedback includes that guest speakers were good and should be used, that Raptor was good but that more graphics should be used and that over half would be interested in taking another CS course and 85% would recommend it to a friend.

Yuen & Robbins [156] present a qualitative case study on five undergraduate biology students who take a CS course (CS0). The course is designed to teach students computing concepts at CT by writing programs in MATLAB and working with data, performing analysis on it. In this "data-driven" context the aim was to better understand the students thought process. They did this through tasks, interviews and they give examples of questions, answers and programming exercises given. They found that students continually go through an organisational process to understand a data sets structure. They also found that visualisation tasks are linked to computational tasks and that they give useful feedback to students when programming.

He et al. [197] present a paper on a teaching method they developed based on CT for a vocational college in China. They use a networking example of how you can create activities that are both engaging and that teach CT concepts. They give an example of this with an experiment, linking each step to a CT concept. For example, the first stage is to "Create a primary domain controller: hb.com" and they link this to Modelling (object construction). During the experiment students, would be guided into thinking not just about how to do each step, but why it was necessary. They found the exam results to be encouraging.

Qualls et al. [198] conducted qualitative and quantitative tests to investigate whether students: 1) Can thoroughly understand the concept of an algorithm, 2) Can effectively apply abstraction to solve a problem and 3) What do students know about efficiency. They tested these questions on a group of students taking an Introduction to Computer Science course which composed both formal lecture and practical lab sessions as well as a variety of assessment. The course covers problem solving strategies with the emphasis being on fundamental programming skills. They found that

- Students understood the concept of an algorithm and saw its importance in CS and that the course had had a positive impact on their performance.
- From interviews, it was clear that students were unable to accurately define abstraction but can illustrate it given examples encountered during the course.
- They found that students seemed to have a generally good understanding of efficiency prior to coming into the course and could transition the idea from a CS concept to one that is applicable in other fields.

Deng et al. [199] give an overview of a method to include CT in CS courses. They first give an overview and some definitions of what CT is. They discuss that using a course as a "carrier" for teaching students CT should do the following:

- Specify the discipline, fundamental problems of the course and abstract its primary thought and technique
- Form a teaching set applicable for the training of subject ideology and methodology
- Choose an appropriate teaching way and method for computational thinking training

Following this they give an example of teaching in a Data Structure and Algorithms class which contains a lot of computing concepts. They suggest a method for teaching which goes from Problem to Model, to Algorithm to Programming.

### 9.6.1 Conclusion

Computational Thinking is a vital skill for $21^{st}$ century workers. Although a lot of research is being conducted into teaching both CT and CS in schools, lots of third level students will never have been exposed to these concepts. It is important that both CS and non-CS students have good problem solving skills and CT can greatly benefit this. Many different methods have been proposed and it seems like a non-compulsory CT course for both CS and non-CS students is a particularly effective and useful method. This requires backing from both administration and teaching staff but the benefits listed both in this section and in Section 7 show that it can be beneficial to all involved. There is also a huge range of ways to teach CT in college contexts, although what most have in common is a more practical, discussion-led courses, and most of these methods seem to be successful. It is thought that, perhaps, CS students will benefit from this as it makes the transition to "traditional programming" easier for them.

## 10 Conclusion

The aim of this paper is to provide educators with an overview of the current research and work that is being done in the area of teaching Computational Thinking. As stated in the introduction, CT is an important skill that we should be teaching students of all ages. To that end the research questions asked were done with the hope that educators from all contexts will be able to find examples of how CT can be incorporated into their classroom. Whether it's through a dedicated CT course/module, an already existing subject or just as a one-off event, CT can be taught in a fun and engaging way whilst teaching students vital skills which can be applied across the curriculum as well as in daily and work life. There is also a variety of different tools and ways in which you can apply them in classrooms, and although more work is needed, there is ways in which CT can be assessed.

[143] Wong, G.K., Cheung, H.Y., Ching, E.C. and Huen, J.M., 2015, December. School perceptions of coding education in K-12: A large scale quantitative study to inform innovative practices. In *Teaching, Assessment, and Learning for Engineering (TALE), 2015 IEEE International Conference on* (pp. 5-10). IEEE.

[144] Black, J., Brodie, J., Curzon, P., Myketiak, C., McOwan, P.W. and Meagher, L.R., 2013, July. Making computing interesting to school students: teachers' perspectives. In *Proceedings of the 18th ACM conference on Innovation and technology in computer science education* (pp. 255-260). ACM.

[145] Zur Bargury, I., 2012, July. A new curriculum for junior-high in computer science. In *Proceedings of the 17th ACM annual conference on Innovation and technology in computer science education* (pp. 204-208). ACM.

[146] Settle, A., Franke, B., Hansen, R., Spaltro, F., Jurisson, C., Rennert-May, C. and Wildeman, B., 2012, July. Infusing computational thinking into the middle-and high-school curriculum. In *Proceedings of the 17th ACM annual conference on Innovation and technology in computer science education* (pp. 22-27). ACM.

[147] Griffin, J., Pirmann, T. and Gray, B., 2016, February. Two Teachers, Two Perspectives on CS Principles. In *Proceedings of the 47th ACM Technical Symposium on Computing Science Education* (pp. 461-466). ACM.

[148] Bargury, I.Z., Muller, O., Haberman, B., Zohar, D., Cohen, A., Levy, D. and Hotoveli, R., 2012, October. Implementing a new computer science curriculum for middle school in Israel. In *Frontiers in Education Conference (FIE), 2012* (pp. 1-6). IEEE.

[149] Morreale, P. and Joiner, D., 2011. Changing perceptions of computer science and computational thinking among high school teachers. *Journal of Computing Sciences in Colleges*, *26*(6), pp.71-77.

[150] Gouws, L.A., Bradshaw, K. and Wentworth, P., 2013, July. Computational thinking in educational activities: an evaluation of the educational game light-bot. In *Proceedings of the 18th ACM conference on Innovation and technology in computer science education* (pp. 10-15). ACM.

[151] Miller, R.B., Kelly, G.N. and Kelly, J.T., 1988. Effects of Logo computer programming experience on problem solving and spatial relations ability. *Contemporary Educational Psychology*, *13*(4), pp.348-357.

[152] Brancaccio, A., Marchisio, M., Palumbo, C., Pardini, C., Patrucco, A. and Zich, R., 2015, July. Problem Posing and Solving: Strategic Italian Key Action to Enhance Teaching and Learning Mathematics and Informatics in the High School. In *Computer Software and Applications Conference (COMPSAC), 2015 IEEE 39th Annual* (Vol. 2, pp. 845-850). IEEE.

[153] Vergara, C.E., Briedis, D., Buch, N., Esfahanian, A.H., Sticklen, J., Urban-Lurain, M., Paquette, L., Dresen, C. and Frazier, K., 2012, October. Work in progress: Integrating computation across engineering curricula: Preliminary impact on students. In *Frontiers in Education Conference (FIE), 2012* (pp. 1-2). IEEE.

[154] Towhidnejad, M., Kestler, C., Jafer, S. and Nicholas, V., 2014, October. Introducing computational thinking through stealth teaching. In *Frontiers in Education Conference (FIE), 2014 IEEE* (pp. 1-7). IEEE.

[155] Shell, D.F., Hazley, M.P., Soh, L.K., Miller, L.D., Chiriacescu, V. and Ingraham, E., 2014, October. Improving learning of computational thinking using computational creativity exercises in a college CSI computer science course for engineers. In *Frontiers in Education Conference (FIE), 2014 IEEE* (pp. 1-7). IEEE.

[156] Yuen, T.T. and Robbins, K.A., 2015. A Qualitative Study of Students' Computational Thinking Skills in a Data-Driven Computing Class. *ACM Transactions on Computing Education (TOCE)*, *14*(4), p.27.

[157] Guo, Y., Wagh, A., Brady, C., Levy, S.T., Horn, M.S. and Wilensky, U., 2016, June. Frogs to Think with: Improving Students' Computational Thinking and Understanding of Evolution in A Code-First Learning Environment. In *Proceedings of the 15th International Conference on Interaction Design and Children* (pp. 246-254). ACM.
54